\documentclass[%
 aps,
 superscriptaddress,
 prb,
 amsmath,amssymb,
 longbibliography,
 reprint,
]{revtex4-2}

\usepackage{dcolumn}
\usepackage{bm}
\usepackage[version=3]{mhchem}
\usepackage[utf8]{inputenc}
\usepackage[T1]{fontenc}
\usepackage{siunitx}

\usepackage{amsmath}
\usepackage{amsfonts}
\usepackage{braket}
\usepackage{siunitx}
\usepackage{dcolumn}
\usepackage{placeins}
\usepackage{booktabs}
\usepackage{graphicx}
\usepackage{dcolumn}
\usepackage{tikz}
\usepackage{array}
\usepackage{xcolor}
\usepackage[table]{xcolor}
\usepackage{xcolor}

\usepackage{paralist,multirow,amssymb}
\usepackage[inline]{enumitem}
\usepackage{algorithm}
\usepackage{algpseudocode}
\usepackage{listings}
\usepackage[percent]{overpic}
\usepackage{bbding}
\usepackage{pifont}
\newcommand{\cmark}{\ding{51}}
\newcommand{\xmark}{\ding{55}}
\newcommand{\tmark}{\ding{115}}

\usepackage{hyperref}
\usepackage[capitalise]{cleveref}
\hypersetup{breaklinks,colorlinks=true,linkcolor=blue,citecolor=blue,urlcolor=blue,filecolor=blue}
\usetikzlibrary{decorations.pathreplacing,calligraphy}
\usetikzlibrary{decorations.markings}
\usetikzlibrary{decorations.pathmorphing}
\usetikzlibrary{shapes.misc}

\tikzset{cross/.style={cross out, draw=black, minimum size=2*(#1-\pgflinewidth), inner sep=0pt, outer sep=0pt},
cross/.default={1pt}}

\definecolor{myblue}{RGB}{0, 102, 204}

\newcommand{\av}[1]{\langle#1\rangle} 

\newcommand{\pos}{\mathbf{r}} 
\newcommand{\wv}{\mathbf{k}} 
\newcommand{\qv}{\mathbf{q}} 
\newcommand{\vep}{\varepsilon} 

\newcommand{\ham}{\hat{H}} 
\newcommand{\dei}[2]{\left(#1|#2\right)} 
\newcommand{\cre}{a^{\dagger}}

\newcommand{\ovl}[1]{\overline{#1}} 

\renewcommand{\d}{\mathrm{d}}

\newcommand{\Harvardc}{\affiliation{Department of Chemistry and Chemical Biology, Harvard University, Cambridge, MA 02138, USA}}

\begin{document}
\title{{\it Ab Initio} Auxiliary-Field Quantum Monte Carlo in the Thermodynamic Limit}

\author{Jinghong Zhang}
\author{Meng-Fu Chen}
\author{Adam Rettig}
\author{Tong Jiang}
\author{Paul J. Robinson}
\author{Hieu Q. Dinh}
\author{Anton Z. Ni}
\author{Joonho Lee}
\email{joonholee@g.harvard.edu}
\Harvardc

\date{\today}

\begin{abstract}
Ab initio auxiliary-field quantum Monte Carlo (AFQMC) is a systematically improvable many-body method, but its application to extended solids has been severely limited by unfavorable computational scaling and memory requirements that obstruct direct access to the thermodynamic and complete-basis-set limits. By combining tensor hypercontraction with $\wv$-point symmetry, we reduce the computational and memory scaling of ab initio AFQMC for solids to $\mathcal O(N^3)$ and $\mathcal O(N^2)$, respectively, with an arbitrary basis, comparable to diffusion Monte Carlo. This enables direct and simultaneous thermodynamic-limit and complete-basis-set AFQMC calculations across insulating, metallic, and strongly correlated solids, without embedding, local approximations, empirical finite-size corrections, or composite schemes. Our results establish AFQMC as a general-purpose, systematically improvable alternative to diffusion Monte Carlo and coupled-cluster methods for predictive ab initio simulations of solids, enabling accurate energies and magnetic observables within a unified framework.
\end{abstract}

\maketitle

\section{Introduction}
Accurate simulation of solid-state systems is crucial across many areas, from fundamental science to technologies spanning disciplines such as condensed matter physics, materials science, and chemistry.
The {\it de facto} workhorse in electronic structure calculations of materials is Kohn-Sham density functional theory (DFT)~\cite{hohenbergInhomogeneous1964,kohnSelfConsistent1965}. While approximate, due to the balance between accuracy and cost ($\mathcal O(N^3)$ with $N$ being the system size), it has been applied to a broad range of solid-state problems~\cite{hasnipDensity2014, jonesDensity2015a,kurthMolecular1999,staroverovTests2004a,csonkaAssessing2009}. The two challenges in DFT have yet to be overcome: strong correlation~\cite{cohenFractional2008,maletStrong2012,cohenInsights2008} and self-interaction error ~\cite{mori-sanchezManyelectron2006}.
Solving these challenges within the DFT framework is an interesting direction, but an alternative active research area is to use many-body methods that go beyond DFT. 
Because the cost of these many-body methods scales more steeply with system size than that of DFT, one often struggles to reach the thermodynamic limit (TDL) and complete basis set (CBS) limit with these methods.
We shall briefly review ongoing efforts in this area and highlight the challenges.

One of the most widely used many-body approaches for solids is fixed-node diffusion Monte Carlo (DMC)~\cite{foulkesQuantum2001,austinQuantum2012,kolorencApplications2011}, which approximately performs imaginary-time evolution. 
It has an attractive $\mathcal O(N^3)$ cost per statistical sample~\cite{williamsonLinearScaling2001} and $\mathcal O(N^2)$ storage cost~\cite{kimQMCPACK2018}. Moreover, it works directly in the CBS limit.
Two sources of bias in DMC are difficult to control and quantify: pseudopotential and fixed-node errors. 
The pseudopotential error in DMC has often been a significant source of error~\cite{mitasNonlocal1991,dzubakQuantitative2017,krogelMagnitude2017}, and a more accurate pseudopotential is still an active research area in DMC~\cite{bennettNew2018,burkatzkiEnergyconsistent2007,casulaLocality2006,drummondTrailNeeds2016}. The fixed-node error is the bias introduced to control the sign problem and maintain statistical efficiency.
While some prior work exists~\cite{nazarovBenchmarking2016,benaliSystematic2020}, quantifying the fixed-node error has been challenging, partly because it operates in the CBS limit, where obtaining exact, reference-theoretical results for comparison is difficult. 

Another popular class of methods applied in a solid-state context is diagrammatic methods, such as the random-phase approximation (RPA)~\cite{bohmCollective1951,bohmCollective1953,pinesCollective1952,gruneisMaking2009,harlAccurate2009} and coupled-cluster (CC) methods~\cite{boothExact2013,gruberApplying2018,yangInitio2014}.
In particular, CC with singles, doubles, and perturbative triples (CCSD(T)) is the gold-standard method for gapped systems with mainly dynamic correlation~\cite{raghavachariFifthorder1989,bartlettCoupledcluster2007,tajtiHEAT2004} at the cost of $\mathcal O(N^7)$ with $\mathcal O(N^4)$ storage~\cite{dattaMassively2021,szaboLinearScaling2023}. 
Unlike DMC, these methods can perform all-electron or projector-augmented-wave calculations, so pseudopotential errors have not been a significant concern.
However, these methods work in a finite basis, so one must extrapolate correlation energies to both the CBS and TDL limits.
While accuracy is reliable, performing CCSD(T) calculations at these limits often requires local correlation approximations, even for simple solids~\cite{Ye2024Oct}, introducing biases stemming from these approximations. 
It is also possible to utilize more advanced size corrections~\cite{gruberApplying2018} and composite corrections via low-level methods~\cite{Ye2024Oct,Neufeld2022Aug,Neufeld2023Oct} to approximate the TDL. The underlying bias of these corrections is often difficult to gauge.

We would also like to mention embedding approaches in which one defines a local impurity problem that is solved accurately, while the rest of the problem is handled by a mean-field method.
Dynamical mean-field theory~\cite{georgesDynamical1996,kotliarStrongly2004,kotliarElectronic2006} and density matrix embedding theory~\cite{kniziaDensityMatrixEmbedding2012,cuiEfficient2020} belong to this category. These approaches help achieve the CBS and TDL for a given impurity method. Still, their accuracy is ultimately limited by the underlying impurity solver and by the locality error (similar in spirit to local correlation methods). Completely removing the locality error is possible by increasing the impurity size, but the (steep) computational scaling of the impurity solver quickly limits this strategy. Because these approaches inevitably introduce locality errors, they motivate alternative methods that can reach the TDL without embedding.

Driven partly by the success of constrained-path auxiliary-field quantum Monte Carlo~\cite{zhangConstrained1995,zhengStripe2017,simonscollaborationonthemany-electronproblemAbsence2020} for the Hubbard model and other related lattice models~\cite{boncaEffects1998,
enjalranConstrainedpath2001,
leeConstrainedpath2021}, {\it ab initio} phaseless AFQMC~\cite{zhangQuantumMonteCarlo2003,mottaInitioComputationsMolecular2018} for studying molecular systems has gained significant attention in recent years~\cite{al-saidiAuxiliaryfield2006,leeUtilizing2020,leeSpectral2021,mahajanTaming2021,mahajanSelected2022,mahajanResponse2023,mahajanCCSDT2025,jiangUnbiasing2025}.
Simulating {\it ab initio} solid-state systems with AFQMC has received relatively less exploration due to the substantially greater computational cost of faithfully performing the TDL and CBS extrapolations without relying on composite schemes~\cite{morales2020accelerate}, downfolding~\cite{maQuantum2015}, or DFT-based size-corrections~\cite{kweeFiniteSize2008,maFinitesize2011, zhangAuxiliaryfield2018,zhangAccurate2023}. 
While relying on these, prior solid-state AFQMC applications have used relatively coarse $\wv$-point sampling~\cite{purwantoPressureinduced2009, Malone2020May, taheridehkordiPhaseless2023}.
This restriction is partly due to the $\mathcal O(N^4)$ computational cost and the $\mathcal O(N^3)$ storage requirements of AFQMC when used with an arbitrary basis, which ultimately constrains the calculations to a smaller basis set and $\wv$-mesh than desired. Compared to the more widely used DMC for the same purpose, AFQMC has indeed seen more limited adoption.

\begin{table*}[htbp]
  \centering
  \caption{Comparison of various electronic structure methods. A \cmark\ denotes a demonstrated capability with published implementations, a \tmark\ indicates a capability that may be theoretically feasible but is currently impractical without additional tricks, and an \xmark\ denotes a capability that is not feasible yet for systems considered in this work. {AFQMC represents our work, noting the status before and after.} 
  DMC and AFQMC costs are per sample. TDL = thermodynamic limit. CBS = complete basis set limit.}
  \label{tab:method-comparison}
  \begin{tabular}{lccccc}
    \toprule
    & DFT       & CCSD      & CCSD(T)   & DMC       & \textbf{AFQMC} \\
    \midrule
    Scaling      & $N^3$              & $N^6$              & $N^7$             & $N^3$              & $N^4\to N^3$      \\
    Metallic     & \cmark             & \cmark             & \xmark            & \cmark             & \cmark      \\
    TDL / CBS    & \cmark             & \tmark         & \tmark        & \cmark             & \tmark{}\textrightarrow\cmark{}   \\
    All-electron & \cmark             & \cmark             & \cmark            & \tmark            & \cmark    \\
    Beyond single-determinant    & \xmark            & \xmark            & \xmark           & \cmark             & \cmark   \\
    \bottomrule
  \end{tabular}
\end{table*}

The purpose of this paper is, by leveraging the new techniques presented here, to place phaseless AFQMC among the above-mentioned state-of-the-art {\it ab initio} techniques and to demonstrate its unique strengths in scalability, storage cost, and flexibility for systematic improvement in studying {\it ab initio} solids. 
Like DMC, AFQMC approximately performs imaginary time evolution with the phaseless constraint~\cite{zhangQuantumMonteCarlo2003,mottaInitioComputationsMolecular2018} to control the sign problem.
Like CC methods, AFQMC can perform all-electron calculations and use a finite basis set, which ultimately requires CBS extrapolation.
The only significant source of error in AFQMC, as long as one can faithfully reach the TDL and CBS limits, is the bias introduced by the phaseless approximation.
With more advanced trial wavefunctions~\cite{al-saidiBond2007,shiSymmetryprojected2014,purwantoAuxiliaryfield2015,sheeAchieving2019,leeUtilizing2020,mahajanSelected2022,mahajanTaming2021,hugginsUnbiasing2022,jiangUnbiasing2025,mahajanCCSDT2025,sukurmaSelfRefinement2025}, one can hope to gauge the level of this bias, even for sophisticated systems, as we will show in this work. Encouragingly, even with a simple Hartree-Fock (HF) trial wave function, AFQMC typically achieves higher accuracy than CCSD~\cite{leeTwentyYearsAuxiliaryField2022}. Moreover, for the same single-determinant trial, AFQMC exhibits substantially smaller bias than DMC~\cite{maloneSystematic2020}.

Our work employs an advanced tensor factorization strategy in quantum chemistry, known as tensor hypercontraction (THC) or interpolative separate density fitting (ISDF)~\cite{parrishExactTensorHypercontraction2013,parrishTensorHypercontractionUniversal2013,luCompression2015,dongInterpolative2018,leeSystematically2020,yehLowScaling2023,yehLowScaling2024}, with $\mathbf k$-point sampling, to bring these costs down to $\mathcal O(N^3)$ compute and $\mathcal O(N^2)$ storage costs, reaching the scaling of DMC.
This development, combined with the efficient use of graphics processing units (GPUs)~\cite{Malone2020May} and newly developed correlation-consistent basis sets~\cite{yeCorrelationConsistent2022}, 
removes the primary algorithmic barrier that previously prevented AFQMC from directly accessing the TDL and CBS simultaneously for the systems considered in this work.

Our work reasserts the competitive advantage of AFQMC over solid-state CC methods through its favorable scaling and flexibility for improvement, while also distinguishing it from DMC through its all-electron capability. Note that although all-electron DMC calculations in solids are, in principle, possible within the full-potential linearized augmented plane wave formalism~\cite{eslerFundamental2010}, they are limited to small cells (< 10 atoms) and light elements due to the intrinsic difficulty of treating core electrons in real space. 
By contrast, AFQMC operates in an orbital-space representation, so all-electron calculations are natural and do not require any modification of the formalism~\cite{purwantoFrozenOrbital2013a}. To make it clear, we listed the computational scaling and limitations of the state-of-the-art electronic structure methods related to solid state applications in Table \ref{tab:method-comparison}.
Hence, this work establishes AFQMC as one of the most promising many-body methods for simulating realistic solid-state systems, as we will elaborate with concrete examples.

This paper is organized as follows: In \cref{sec:theory}, we briefly overview the theory for the conventional AFQMC algorithm applied to periodic systems, which has been presented in ref.~\cite{Malone2020May,mottaHamiltonian2019}; the basic notations and formalism of periodic THC using ISDF~\cite{yehLowScaling2023}, and AFQMC combined with THC for systems with $\wv$-point symmetry. 
We also analyze the sources of error in the calculations and systematically correct them using a consistent protocol. In \cref{sec:results}, we report the cohesive energies for several representative simple solids calculated using AFQMC: diamond, silicon, BCC lithium, and FCC aluminum, as well as the Heisenberg exchange constant, $J$, for strongly correlated systems, NiO and \ce{CaCuO2}. 
More relevant details, including the theoretical aspects of our method, our extrapolation scheme, and other technical details, are provided in the appendices. 

\section{Theory}\label{sec:theory}

\subsection{Overview of AFQMC}
We briefly summarize standard AFQMC to establish notation and highlight the computational bottlenecks.
In particular, we will focus on its application to solid-state systems under periodic boundary conditions. 
Some low-level AFQMC details will be omitted for clarity, but we present them in \cref{app:afqmc} for completeness.

In the Bloch orbital basis, the second-quantized electronic Hamiltonian reads
\begin{equation}
\begin{aligned}
        \ham &= \sum_{pq, \wv}\sum_{\sigma} h^\wv_{pq}\cre_{p\wv\sigma} a_{q\wv\sigma} \\
        &+ \frac{1}{2}\sum_{pqrs}\sum_{\wv \wv' \mathbf{q}}\sum_{\sigma, \sigma'} \dei{p\wv ,r\wv+ \mathbf{q}}{q\wv' + \mathbf{q},s\wv'}\\
        &\qquad\cre_{p\wv \sigma}\cre_{q\wv'+\mathbf{q}\sigma'}a_{s\wv'\sigma'}a_{r\wv+ \mathbf{q}\sigma},
    \label{ham}
\end{aligned}
\end{equation}
where $h^\wv_{pq}$ and $\dei{p\wv ,r\wv+ \mathbf{q}}{q\wv' + \mathbf{q},s\wv'}$ are the one-electron and two-electron integrals, respectively, and $\mathbf k$, $\mathbf k'$, and $\mathbf q$ are sampled from the first Brillouin zone. 
Since the Coulomb kernel is positive definite, we can perform the Cholesky decomposition on the two-electron integral:
\begin{equation}
    \dei{p\wv ,r\wv+ \mathbf{q}}{q\wv' + \mathbf{q},s\wv'} = \sum_{\gamma}L_{p\wv , r\wv+ \mathbf{q}}^{\gamma, \mathbf{q}}L_{s\wv',q\wv' + \mathbf{q}}^{\gamma, \mathbf{q}*}, 
    \label{chol}
\end{equation}
where $\mathbf L$ represents the Cholesky vectors and this is the cause of $\mathcal O(N^3)$ storage cost in solid-state AFQMC (more precisely $\mathcal O (N_k^2 n^3)$ with $N_k$ being the number of $\mathbf k$-points and $n$ being the size of the unit cell).

For AFQMC, we write the two-body part of the Hamiltonian into the sum of operator squares to which we apply the Hubbard-Stratonovich transformation~\cite{hubbard1959calculation,stratonovich1957method}. By defining
\begin{equation}
    \hat{H}_1' = \sum_{pq,\wv}\sum_{\sigma}\left[ h^\wv_{pq} - \frac{1}{2}\sum_{r\mathbf{q}} L^{\gamma, \mathbf{q}}_{p\wv, r\wv + \mathbf{q}}L^{\gamma, \mathbf{q}*}_{ q\wv,r\wv + \mathbf{q}} \right]\cre_{p\wv\sigma} a_{q\wv\sigma},
    \label{h1emod}
\end{equation}
the remaining two-body part can be written as
\begin{equation}
    \ham_2' = \frac{1}{2}\sum_{\gamma, \qv} \hat{L}_{\gamma, \qv}\hat{L}^\dagger_{\gamma, \qv},
\end{equation}
where 
\begin{equation}
    \hat{L}_{\gamma, \mathbf{q}} = \sum_{pr, \wv}\sum_{\sigma}L^{\gamma, \mathbf{q}}_{p\wv , r\wv+ \mathbf{q}}\cre_{p\wv \sigma}a_{r\wv+ \mathbf{q}, \sigma}.
    \label{Lgq}
\end{equation}
We can further write the two-body Hamiltonian as
\begin{equation}
    \ham_2' = -\frac{1}{2}\sum_{\gamma\mathbf{q}} ((\hat{L}^{+}_{\gamma, \mathbf{q}})^2 + (\hat{L}^{-}_{\gamma, \mathbf{q}})^2),
    \label{eq:h2p}
\end{equation}
where the $\hat{L}^{\pm}_{\gamma, \mathbf{q}}$ operators are defined as
\begin{equation}
    \hat{L}^{\pm}_{\gamma, \mathbf{q}} := c_{\pm}\left(\frac{\hat{L}_{\gamma, \mathbf{q}} \pm \hat{L}^\dagger_{\gamma, \mathbf{q}}}{2}\right),
    \label{Lpm}
\end{equation}
with coefficients $c_+ = i$ and $c_- = 1$, achieving the necessary sum-of-squares form.

AFQMC approximates the exact ground state, $\ket{\Psi_0}$, of this Hamiltonian by performing an imaginary time evolution on an initial state $\ket{\Psi_I}$ which is not orthogonal to the ground state, (with discrete times)~\cite{mottaInitioComputationsMolecular2018},
\begin{equation}
    \ket{\Psi_0} \propto \lim_{N \to \infty} \left(\mathrm{e}^{-\Delta \tau\hat{H}}\right)^N\ket{\Psi_I}.
    \label{ite}
\end{equation}
Using the Hubbard-Stratonovich transformation~\cite{hubbard1959calculation,stratonovich1957method}, we can write the short-time propagator as an integral over the auxiliary fields
\begin{equation}
    \mathrm{e}^{-\Delta \tau \hat{H}} = \int \mathrm{d}^{N_{\text{aux}}} \mathbf{x}\ p(\mathbf{x}) \hat{B}(\mathbf{x}) + \mathcal{O}(\Delta\tau^2),
    \label{eq: HS}
\end{equation}
where $N_{\mathrm{aux}} = N_\gamma N_\qv$, $p(\mathbf{x})$ is the standard Gaussian distribution, and $\hat{B}(\mathbf{x})$ is the propagator given by
\begin{equation}
    \hat{B}(\mathbf{x}) = e^{-\Delta \tau \ham_1'/2}e^{\sqrt{\Delta\tau}\sum_{\gamma, \mathbf{q}}
     (x^+_{\gamma, \mathbf{q}} \hat{L}^+_{\gamma, \mathbf{q}} + x^-_{\gamma, \mathbf{q}} \hat{L}^-_{\gamma, \mathbf{q}})} e^{-\Delta \tau \ham_1'/2}.
\end{equation}
We sample these propagators and represent the global wavefunction at a given imaginary time as (without importance sampling),
\begin{equation}
    |\Psi(\tau)\rangle = \sum_\alpha w_\alpha 
    |\Phi_\alpha\rangle,
\end{equation}
where $|\Phi_\alpha\rangle$ is the $\alpha$-th walker wavefunction (i.e., a Slater determinant) and $w_\alpha$ is the corresponding weight.
In practice, we control the variance using importance sampling techniques, specifically the force bias and mean-field shift, as well as the phaseless approximation
The importance sampling and phaseless constraints are achieved by introducing a trial wavefunction, $|\Psi_T\rangle$, whose quality ultimately determines the accuracy of phaseless AFQMC.
We will discuss the phaseless error in \cref{erroranalysis}.
More rigorous implementation details are provided in Appendix~\ref{app:mfshift}.

According to the Thouless theorem~\cite{thoulessStability1960,thoulessVibrational1961}, the action of $\hat{B}(\mathbf{x})$ on a Slater determinant parametrized by $\mathbf{C}$ is given by the matrix
\begin{equation}
    \mathbf{C}' = \mathbf{B}(\mathbf{x})\mathbf{C}.
\end{equation}
We define the one-body Hamiltonian matrix as 
\begin{equation}
    [h_1]_{p\wv, q\wv'} = \left( h^\wv_{pq} - \frac{1}{2}\sum_{r\qv} L^{\gamma, \mathbf{q}}_{p\wv, r\wv + \mathbf{q}}L^{\gamma, \mathbf{q}*}_{ q\wv,r\wv + \mathbf{q}}\right) \delta_{\wv, \wv'},
\end{equation}
and HS transformed two-body propagation matrix $\mathbf{V}_{\mathrm{HS}}$ as
\begin{equation}
    [V_{\mathrm{HS}}]_{p\wv, q\wv'} = \sum_{\gamma} (x_{\gamma, \wv'-\wv}L^{\gamma, \wv'-\wv}_{p\wv , q\wv'}  + y_{\gamma, \wv - \wv'} L^{\gamma, \wv-\wv'*}_{ q\wv', p\wv}),
\end{equation}
where
\begin{equation}
    x_{\gamma, \qv} = \frac{1}{2}(ix^+_{\gamma,\qv} + x^-_{\gamma, \qv}) ; \quad y_{\gamma, \qv} = \frac{1}{2}(ix^+_{\gamma,\qv} - x^-_{\gamma, \qv}).
    \label{auxfields}
\end{equation}
Then, the propagator matrix $\mathbf{B}(\mathbf{x})$ can be written as
\begin{equation}
    \mathbf{B}(\mathbf{x}) = e^{-\Delta\tau \mathbf{h}_1/2}e^{\sqrt{\Delta \tau} \mathbf{V}_{\mathrm{HS}}}e^{-\Delta\tau \mathbf{h}_1/2},
\end{equation}
from which we can see the propagation step in the original AFQMC formalism scales as $\mathcal{O}(N_k^3 n^3)$, which does not exhibit any scaling reduction from the $\wv$-point symmetry. 
This completes the summary of the relevant technical details for walker propagation in AFQMC.
The computation of estimators and the standard Cholesky-based local energy computation are summarized in \cref{app:estimators}.

\subsection{Periodic THC using ISDF}
Periodic ISDF~\cite{hennekeFast2020,maRealizing2021,rettigEven2023,yehLowScaling2023} aims to decompose the orbital product into the following form:
\begin{equation}
    \rho^{\qv, \wv}_{\mu\nu}(\pos) := \phi^{\wv *}_\mu (\pos) \phi_\nu^{\wv+ \mathbf{q}}(\pos) = \sum_P \zeta_P^{\qv}(\pos) \phi^{\wv *}_\mu (\pos_P) \phi_\nu^{\wv+ \mathbf{q}}(\pos_P),
\end{equation}
where the set of interpolating points $\{\pos_P\}$ is a subset of all grid points $\{\pos\}$, and $\{\zeta_P^\qv(\pos)\}$ are the interpolating vectors. Since the orbitals satisfy Bloch's theorem, it is straightforward to verify that the interpolating vectors $\{\zeta_P^\qv(\pos)\}$ are also Bloch functions. By defining 
\begin{equation}
    M_{PQ}^\qv := \int\d\pos \d \pos' \zeta_P^{\qv}(\pos) \frac{1}{|\pos - \pos'|}\zeta_Q^{\qv*}(\pos'),
\end{equation}
we can bring the electron repulsion integral (ERI) tensor into the following THC form

\begin{equation}
\begin{aligned}
    (\mu\wv, \nu\wv+\mathbf{q}|\lambda\wv' + \mathbf{q} , \sigma\wv') &= \sum_{PQ}\phi^{\wv*}_\mu(\pos_P)\phi^{\wv+\qv}_\nu(\pos_P)M_{PQ}^\qv\\
    &\phi^{\wv'+\qv*}_\lambda(\pos_Q)\phi^{\wv'}_\sigma(\pos_Q),
\end{aligned}
\end{equation}
where $P$ and $Q$ label ISDF interpolating points, and the total number of interpolating points $N_{\mathrm{ISDF}}$ needed to achieve a fixed accuracy is found to be a constant multiple of the number of basis functions $M$, namely $N_{\mathrm{ISDF}} = c_{\mathrm{ISDF}} M$ with $c_{\mathrm{ISDF}} \approx 10$--$20$~\cite{malone2018overcoming,leeSystematically2020,yehLowScaling2023} (see also our own convergence test in Appendix \ref{app:thcerror}).
We follow the procedure of Ref.~\cite{ matthewsImproved2020} to select the interpolating points. Defining
\begin{equation}
\mathbf{S}^\mathbf{q} := \bm{\rho}^{\mathbf{q}\dagger}\bm{\rho}^\mathbf{q},
\end{equation}
with $\bm{\rho}^\qv=\rho^\qv_{\wv ij,\pos}$ of dimension $(N_k M^2)\times N_{\rm grid}$, we then apply a pivoted Cholesky decomposition to $\mathbf{S}^\mathbf{q}$. During this iterative factorization, the algorithm selects interpolating points until the prescribed threshold $\epsilon_{\mathrm{ISDF}}$ is reached. To avoid $N_k$ repetitions of the Cholesky procedure, we take the interpolating points only at $\qv = \mathbf{0}$ and use the same set of grid points for all other $\wv$ points. This has been shown to introduce negligible errors in the final result in Ref.~\cite{yehLowScaling2023}. 

\subsection{k-point THC-AFQMC}
THC-AFQMC without $\wv$-point symmetry was first reported in Ref.~\citenum{malone2018overcoming}, so we will focus here on key differences due to the use of $\wv$-point symmetry in our work.
We refer to this THC approach with $\wv$-point symmetry as $\wv$-THC-AFQMC.

For each $\qv$, 
$M_{PQ}^{\qv}$ is a positive definite Hermitian matrix, which can be derived from the positive definiteness of the Coulomb kernel $1/r_{12}$. Hence, we can perform Cholesky decomposition on it, 
\begin{equation}
    M^{\qv}_{PQ} = \sum_{\gamma}R^\qv_{P\gamma}R^{\qv*}_{Q \gamma}.
\end{equation}
In this factorization, we do not exploit the low-rankness since $M^{\qv}_{PQ}$ is already rank-revealed, so we retain the full-rank of $M^{\qv}_{PQ}$. 
By defining
\begin{equation}
\hat{L}_{\gamma, \qv} \equiv \sum_{P\wv}\sum_{pr\sigma}\psi_p^\wv(\pos_P)^* \psi_r^{\wv + \mathbf{q}}(\pos_P)R^\qv_{P\gamma}\cre_{p\wv\sigma}a_{r\wv + \mathbf{q}\sigma},
\end{equation}
where $\psi_p^\wv(\pos_P)$ denotes the orthogonalized basis functions evaluated on the ISDF grid, we recover the same two-body Hamiltonian as in \eqref{eq:h2p}. With this form, the contraction in the two-body propagator can be performed differently from that of conventional algorithms to reduce the scaling. 

The key step in propagation is the action of the matrix $\mathbf{V}_{\mathrm{HS}}$ on the walker wavefunction, which can be written as
\begin{equation}
\begin{aligned}
    [\hat{V}_{\mathrm{HS}}\Phi]_{p\wv, i\wv'} &= \sum_{r\wv'', P}([Rx]^{\wv'' - \wv}_P + [R^*y]^{\wv - \wv''}_P)\\
 &\psi^{\wv*}_{p}(\pos_P)\psi^{\wv''}_{r}(\pos_P)[\Phi]_{r\wv'', i\wv'},
\end{aligned}
\end{equation}
where $ [Rx]_P^\qv = \sum_{\gamma} R^\qv_{P\gamma}\, x_{\gamma,\qv} $, $ [R^*y]_P^\qv = \sum_{\gamma} R^{\qv*}_{P\gamma}\, y_{\gamma,\qv} $ and $x_{\gamma, \qv}$, $y_{\gamma, \qv}$ are the auxiliary fields defined in equation \eqref{auxfields}. 
This contraction strategy avoids the explicit construction of the two-body propagator and is key to reducing the memory footprint and computational cost.
Pre-contracting the Bloch basis functions with the walker wavefunction reduces the cost from $\mathcal{O}(N_k^2 M^2 ( N_{\mathrm{ISDF}} + N_k n_{\mathrm{occ}}))$ (building and applying $\mathbf{V}_{\mathrm{HS}}$~\cite{malone2018overcoming}) to $\mathcal{O}(N_k^2 N_{\mathrm{ISDF}} n_{\mathrm{occ}} (N_k + M))$, as shown in \cref{algorithm:vhs}, where
$n_{\mathrm {occ}}$ is the number of occupied orbitals. 
More details are given in Appendix~\ref{propdetails}.

The local energy expression with THC-decomposed ERI is then given by
\begin{equation}
    \begin{aligned}
    E_{L,\alpha} &= \frac{\braket{\Psi_T|\hat{H}|\Phi_\alpha}}{\braket{\Psi_T|\Phi_\alpha}}\\
    &= \sum_{pq\wv, \sigma} h^\wv_{pq} G^{\alpha,\sigma}_{p\wv, q\wv}\\
    &+ \sum_{\substack{pqrs, PQ,\wv\wv'\qv\\\sigma,\sigma'}}\psi^{\wv*}_p(\pos_P)\psi^{\wv+\qv}_r(\pos_P)M_{PQ}^\qv\\
    &\psi^{\wv'+\qv*}_q(\pos_Q)\psi^{\wv'}_s(\pos_Q)(G^{\alpha,\sigma}_{p\wv, r\wv+\qv }G^{\alpha,\sigma'}_{q\wv' + \qv,s\wv'} \\
    &- \delta_{\sigma\sigma'}G^{\alpha,\sigma}_{p\wv, s\wv' }G^{\alpha,\sigma'}_{q\wv' + \qv,r\wv+\qv}),
    \label{eloc_isdf}
\end{aligned}
\end{equation}
where $\ket{\Psi_T}$ is the trial wavefunction, and
\begin{equation}
    G^{\alpha,\sigma}_{p\wv, r\wv+\qv} = \frac{\braket{\Psi_T|a_{p\wv\sigma}^\dagger a_{r\wv+\qv\sigma}|\Phi_\alpha}}{\braket{\Psi_T|\Phi_\alpha}}
\end{equation}
is the one-particle Green's function calculated with the $\alpha$-th walker.
With appropriate contraction paths, the scaling for evaluating the local energy is $\mathcal{O}(N_k^3n^3)$, or $\mathcal{O}(N_k^2n^4)$, depending on the choice of algorithm. 
This is better than the conventional scaling of $\mathcal O(N_k^3n^4)$. More theoretical details are given in \cref{app:estimators}.

In principle, it is possible to achieve the same scalings by using a plane-wave basis set as a primary basis~\cite{zhangQuantumMonteCarlo2003, suewattanaPhaseless2007}. 
When combined with the projector augmented-wave method~\cite{taheridehkordiPhaseless2023}, pseudopotential errors can be further reduced. However, plane waves do not provide a compact representation for recovering correlation energy~\cite{purwantoFrozenOrbital2013a}, which makes CBS extrapolation particularly challenging. In contrast, our $\wv$-THC-AFQMC formulation works in a much more compact Gaussian orbital basis, enabling CBS extrapolation in the same efficient manner as in standard quantum chemistry calculations~\cite{helgakerBasisset1997,halkierBasisset1998}.

\subsection{Sources of the error }\label{erroranalysis}
Before moving to the results and discussion, we provide a complete list of sources of error in our AFQMC calculations along with the strategies we employed to assess and mitigate them. 
We assume that the THC factorization, time step, population control bias, finite-size effects, and basis-set incompleteness errors are negligible due to our appropriate handling.

For the purpose of the current study, three remaining, relatively significant errors come from: 
\begin{enumerate}
    \item Atomic phaseless error. It is only relevant in the cohesive energy calculations, which is from the phaseless constraint imposed in the isolated-atom reference calculations.
    \item Crystalline phaseless error. It is associated with the phaseless bias introduced in solid-state systems calculations, which depends on the quality of the trial wavefunction. We use spin-restricted HF (RHF) or spin-unrestricted (UHF) trial wavefunctions in this work. This trial choice is the most scalable for AFQMC.
    \item Pseudopotential error. In AFQMC, this is an optional source of error, defined as the difference between the relative energy obtained with a pseudopotential (using the corresponding pseudopotential basis set) and the all-electron relative energy calculated with Dunning's correlation-consistent basis sets. 
\end{enumerate}
Each of these errors can affect the computed relative energies by several m$E_h$, and their relative importance depends on problems.

To quantify the atomic phaseless error, near-exact atomic reference energies could be obtained with brute-force methods such as full configuration interaction (FCI) or selected CI ~\cite{huronIterative1973,ginerUsing2013,
holmesHeatBath2016,
schriberCommunication2016,
sharmaSemistochastic2017,
tubmanModern2020}. 
To obtain accurate cohesive energies, we evaluate the energy of the isolated atom surrounded by ghost atoms in order to account for the basis set superposition error (BSSE)~\cite{boysCalculation1970,maschioPeriodic2010,Malone2020May}. 
This setup substantially increases the number of basis functions, making exact solvers computationally demanding if not infeasible. 
Consequently, we compute the atomic correction without ghost atoms and add it to the AFQMC results obtained with ghost atoms, assuming that ghost atoms introduce only a constant shift in the total energy.

To assess the crystalline phaseless error, which depends on the $\wv$-point mesh, the ideal approach would be to perform exact calculations in the thermodynamic and complete basis set limits. Since such calculations are not tractable at the system size scale needed for the present study, we instead estimate the error using AFQMC with Configuration Interaction with Singles and Doubles (CISD) trial wavefunction (CISD-AFQMC)~\cite{mahajanCCSDT2025} on smaller cells with a small basis set. We then assume that the crystalline phaseless error does not increase significantly with system size and varies little with the basis set size. 

Finally, to estimate the pseudopotential error, we compute the difference between pseudopotential-based relative energies and corresponding all-electron values. Due to the limitations of available all-electron SCF solvers, however, these calculations can only be performed for small supercells in a double-zeta basis, since convergence becomes increasingly difficult for larger cells and basis sets because of linear dependency issues.

Certainly, the above corrections for the crystalline phaseless error, as well as for the pseudopotential errors, are not exact, since they are evaluated only at the DZ level with relatively small supercells. For the crystalline phaseless error, this approximation is unavoidable because exact solvers are computationally infeasible for large basis sets and supercells, and the correction can only be reliably estimated at lower levels of theory. Nevertheless, we observe only minor changes in the pseudopotential corrections when increasing the $\wv$-mesh from $2\times2\times2$ to $3\times3\times3$ (see Appendix \ref{detailerroranalysis} for details), suggesting that the residual error in these corrections is unlikely to exceed their magnitudes. 
Ultimately, pseudopotential errors can be completely eliminated by utilizing all-electron calculations. We will report more all-electron results in the future by leveraging recent advances~\cite{leeSystematically2020,dinhEfficient2025}. 

In what follows, we analyze the impact of each error source in detail for each system and, where possible, apply corrections to reduce its effect on the final results. 

\section{Results}\label{sec:results}
In this section, we present benchmark calculations obtained with our $\wv$-THC-AFQMC approach for both structural and magnetic properties across representative classes of solids. 
We provide an overview of the systems studied in this work and the corresponding system sizes in Table~\ref{tab:syssizes}.
As a comparison, we also provide a summary of prior solid-state AFQMC calculations in \cref{tab:priorafqmc}.
In what follows, we refer to $\wv$-THC-AFQMC as AFQMC for simplicity, unless the context is insufficient to make it unclear.

We first report cohesive energies for prototypical semiconductors (diamond and silicon) and simple metals (BCC Li and FCC Al), with careful attention to thermodynamic-limit extrapolation. We then turn to correlated transition-metal oxides, where we extract magnetic exchange couplings for NiO and CaCuO$_2$ from total-energy differences. Furthermore, we report the local magnetic moment in the antiferromagnetic (AFM) ground state of NiO. 

Together, these results assess the accuracy, robustness, and practical scope of the method for insulating, metallic, and strongly correlated materials. 
Achieving reliable TDL and CBS limit estimates generally requires simulations on dense $\wv$ meshes and large basis sets, which was not possible without the techniques developed in this work. 

\begin{table}[htbp]
    \centering
    \caption{$\wv$-mesh, basis set, total number of orbitals ($N_\mathrm{o}$), and total number of electrons ($N_\mathrm{e}$) for the largest calculation performed for each system considered in this work. Note that for NiO and \ce{CaCuO2}, the computational cost is approximately doubled relative to the other systems because we use UHF (instead of RHF) trial wavefunctions. 
    }
    \begin{tabular}{ccccc}
        \toprule
        System & $\wv$-mesh & basis set & $N_\mathrm{o}$ & $N_\mathrm{e}$\\
        \midrule
        C & $5\times 5\times 5$ & QZ & 13000 & 1000\\
        Si & $5\times 5\times 5$ & QZ & 13000 & 1000\\
        Li & $5\times 5\times 5$ & QZ & 9250 & 750\\
        Al & $4\times 4\times 4$ & QZ & 12032 & 768\\
        \multirow{2}{*}{NiO} & $4\times 4\times 4$ & DZ & 6016 & 3072\\
         & $3\times 3\times 3$ & QZ& 8046 & 1296 \\
        \ce{CaCuO2} & $3\times 3\times 2$ & DZ & 5904 & 2952\\
        \bottomrule
    \end{tabular}
    \label{tab:syssizes}
\end{table}
\subsection{Semiconductors}

The ground state of prototypical semiconductors is relatively well established as a benchmark for various density functionals and many-body approaches~\cite{shishkinSelfconsistent2007,lejaeghereReproducibility2016}. 
Here, we study them to validate our AFQMC approach and to stress-test the computational cost required for reliable basis-set and size extrapolations.

\subsubsection{Diamond}
\begin{figure}
    \centering
    \includegraphics[width=1.0\linewidth]{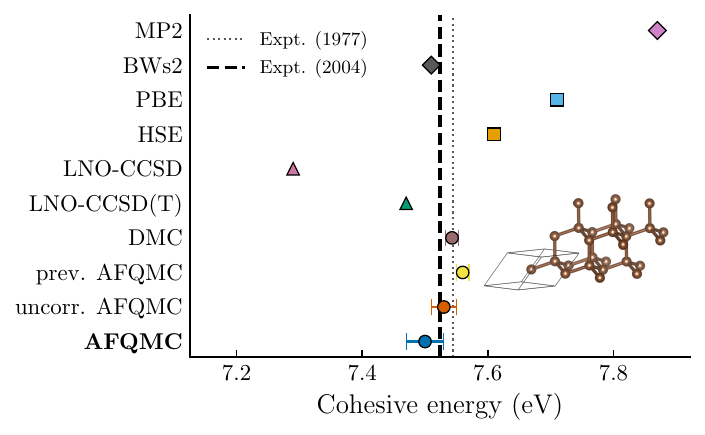}
    \caption{Error in the cohesive energy of diamond crystal predicted by different methods relative to experimental values~\cite{brewerCOHESIVE1977, Schimka2011Jan,ruscicIntroduction2004}. 
MP2, LNO-CCSD, and LNO-CCSD(T) values are taken from Ref.~\cite{Ye2024Oct}; The size-consistent Brillouin–Wigner approach (BWs2) value is taken from~\cite{chenRegularized2025}; PBE and HSE correspond 
to the Heyd-Scuseria-Ernzerhof and Perdew-Burke-Ernzerhof functionals, respectively,  
with values from Ref.~\cite{Schimka2011Jan}; DMC result is from Ref.~\cite{benaliSystematic2020}. The previous AFQMC result (denoted as ``prev. AFQMC'') is taken from Ref.~\cite{Malone2020May}, the AFQMC result is from this work, which is corrected for atomic phaseless, crystalline phaseless and pseudopotential errors as described in the main text. The uncorrected AFQMC value is denoted as uncorr. AFQMC in the plot and is also given in Appendix \ref{cohdata}.
}
    \label{fig:diamonderr}
\end{figure}

We first apply our method to diamond, one of the most-studied systems. It is considered a single-reference system, as manifested by previous DFT~\cite{Schimka2011Jan}, coupled cluster~\cite{Ye2024Oct}, single-determinant DMC~\cite{Benali2020Nov} and AFQMC~\cite{Malone2020May} calculations.
DFT methods, PBE and HSE, overestimate the cohesive energy by 1-2 \%, as shown in Figure \ref{fig:diamonderr}. 
{Previous DMC and AFQMC also match the experiment closely. We note that the prior AFQMC calculation~\cite{Malone2020May} used TZ $3\times3\times3$ and $4\times 4\times 4$ $\wv$-point calculations to obtain the TDL correction for QZ. We verified that this error is small via direct TDL extrapolations of the QZ basis set in our calculations, as shown in \cref{fig:tdlextrap}(a) in \cref{app:extraptdl}.} MP2 substantially overestimates the cohesive energy, whereas the size-consistent Brillouin-Wigner perturbation theory (BWs2)~\cite{carter2023repartitioned, chenRegularized2025} yields a much more accurate result. CCSD performs worse than AFQMC, consistent with observations in previous studies~\cite{leeTwentyYearsAuxiliaryField2022}. As expected, CCSD(T) significantly improves upon CCSD.

 Using AFQMC, we obtain a cohesive energy of 7.53(2) eV, in close agreement with the experimental values of 7.55 eV~\cite{brewerCOHESIVE1977, Schimka2011Jan} and 7.52 eV~\cite{ruscicIntroduction2004} (the latter is calculated from the formation enthalpy difference between two species at 0 K in the ATcT table and applying the zero point energy (ZPE) correction from Ref.~\cite{Schimka2011Jan}, and the error bar is negligible). Ours is the first AFQMC study to directly target the TDL and CBS simultaneously, without any local approximations or composite corrections, thereby enabling transparent analysis of finite-size and basis-set effects. 

One may wonder whether AFQMC can be used to obtain finite-size and basis-set corrections for other many-body methods, such as coupled-cluster methods. By defining
\begin{equation}
    \Delta E_{\mathrm{coh}}(\mathrm{TDL}) = E_{\mathrm{coh}}(\mathrm{TZ};\mathrm{TDL}) - E_{\mathrm{coh}}(\mathrm{TZ}; 3\times3\times3)
\end{equation}
\begin{equation}
    \Delta E_{\mathrm{coh}}(\mathrm{CBS}) = E_{\mathrm{coh}}(\mathrm{CBS};\mathrm{TDL}) - E_{\mathrm{coh}}(\mathrm{TZ}; \mathrm{TDL})
\end{equation}
we find $\Delta E_{\rm coh}^{\rm AFQMC}({\rm TDL})=0.26(2)$ eV and $\Delta E_{\rm coh}^{\rm AFQMC}({\rm CBS})=0.22(2)$ eV. For comparison, MP2 yields $\Delta E_{\rm coh}^{\rm MP2}({\rm TDL})=0.21$ eV and $\Delta E_{\rm coh}^{\rm MP2}({\rm CBS})=0.23$ eV~\cite{Ye2024Oct}. Applying our AFQMC finite size and basis set corrections to LNO-CCSD and LNO-CCSD(T) shifts their cohesive energies to 7.34(2) eV and 7.52(2) eV, respectively, bringing both into slightly better agreement with experiment. While the most reliable approach is to perform TDL extrapolation consistently within the same method, AFQMC may yield more accurate correction terms than other low-scaling methods, such as MP2 or RPA.

\begin{figure}
    \centering
    \includegraphics[width=0.8\linewidth]{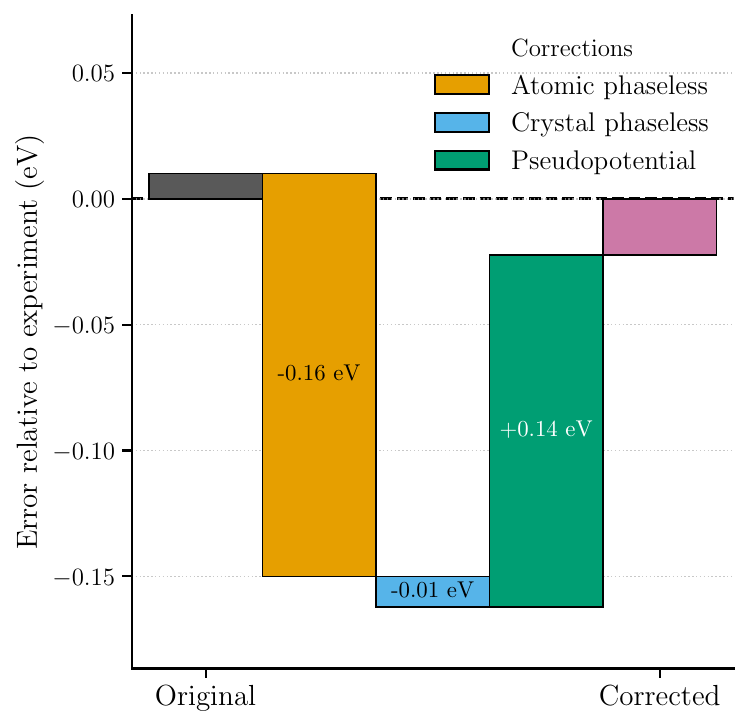}
    \caption{Errors in AFQMC from different sources contributing to the cohesive energy of diamond, including the atomic phaseless error, crystalline phaseless error, and pseudopotential error. The atomic phaseless error is corrected using semistochastic heat-bath configuration interaction (SHCI) for TZ and QZ basis sets, and the results are extrapolated to the complete basis set (CBS) limit. The crystalline phaseless error is corrected using CISD-AFQMC on a $2\times2\times2$ supercell in the DZ basis. The pseudopotential error was corrected using a $3\times3\times3$ supercell at the DZ level. For clarity, the error bars for the corrections are not shown, as each is smaller than 0.01~eV, which would increase the final uncertainty by 0.01~eV.
}
    \label{fig:erranalysis_C}
\end{figure}

How can we understand the near-exact accuracy of AFQMC presented here?
As first reported in Ref.~\cite{leeTwentyYearsAuxiliaryField2022}, AFQMC performs poorly for atomic ground states with total orbital angular momentum other than $S$. For example, the atomic energy error for carbon is 3.74~kcal/mol in the aug-cc-pVTZ basis set, corresponding to a cohesive-energy error of approximately 0.16~eV.
The apparent agreement observed here must therefore be the result of a substantial cancellation between the atomic phaseless error and other error sources. To investigate this behavior, we apply our error-analysis strategy described in Sec.~\ref{erroranalysis} to this system. The corresponding results are shown in Fig. \ref{fig:erranalysis_C}.

The crystalline phaseless error is small, which is expected for a single-reference problem like diamond. The other main error in diamond is the pseudopotential error, which nearly cancels the atomic phaseless error. After correcting all these errors, the error of the cohesive energy of diamond becomes $\sim$0.03 eV compared to the experimental value, which is within the error bar of the corrected values. Interestingly, AFQMC with a single-determinant trial wavefunction becomes slightly lower in energy than CISD-AFQMC as we simulate larger supercells. To validate this, we calculated the energies for various anisotropic supercells, and we observed the same trend (see Appendix \ref{detailerroranalysis} for details).
Assuming CISD-AFQMC is nearly exact compared to FCI or selected CI (which has been verified for smaller supercells) for these supercell sizes, we conclude that single-determinant AFQMC could show slight non-variationality in larger supercells for diamond. 
\subsubsection{Silicon}
Having established the reliability of our method in diamond, we now turn to silicon as the next system of interest. The shared crystal structure enables a straightforward comparison. Silicon’s longer bond length and weaker covalent bonding character make it a complementary and important benchmark for AFQMC. 
{Moreover, the band gap of silicon (1.12 eV~\cite{strehlowCompilation1973}) is much smaller than that of diamond (5.5 eV~\cite{strehlowCompilation1973}). Therefore, electron correlation effects are expected to become more pronounced in silicon.} 

\begin{figure}
    \centering
    \includegraphics[width=1.0\linewidth]{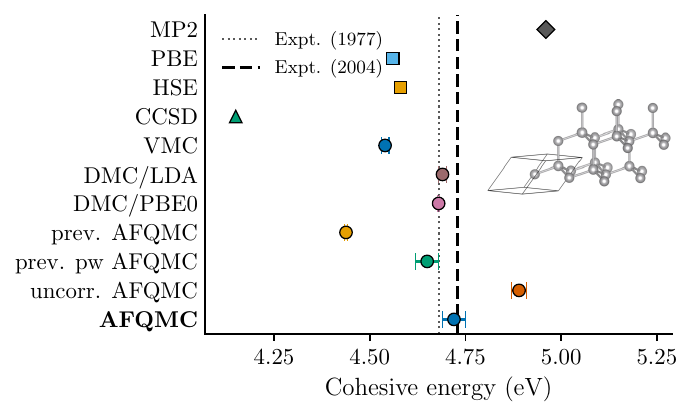}
    \caption{
    Error in the cohesive energy of silicon crystal predicted by different methods relative to experiment~\cite{brewerCOHESIVE1977, Schimka2011Jan,ruscicIntroduction2004}. 
The MP2 value and the CCSD value are from Ref.~\cite{McClain2017Mar}; PBE and HSE values are taken from Ref.~\cite{Schimka2011Jan}; VMC and DMC/LDA results are from Ref.~\cite{leungCalculations1999}, and the DMC/PBE0 result is from Ref.~\cite{Annaberdiyev2021May}; prev. AFQMC denotes the previous AFQMC result from Ref.~\cite{morales2020accelerate}; prev. pw AFQMC denotes the previous plane-wave AFQMC result from~\cite{zhangQuantumMonteCarlo2003}. Note that we apply the ZPE correction to the experimental value, so we removed it from the pw AFQMC result. AFQMC result is from this work, which is corrected for atomic phaseless, crystalline phaseless and pseudopotential errors as described in the main text. The uncorrected AFQMC value is denoted as uncorr. AFQMC in the plot and is also given in Appendix \ref{cohdata}.
}
    \label{fig:siliconerr}
\end{figure}

In \cref{fig:siliconerr}, we present cohesive energies of silicon computed by various approaches.
The experimental values are 4.68 eV~\cite{brewerCOHESIVE1977, Schimka2011Jan} and 4.73 eV~\cite{ruscicIntroduction2004}, where the latter value was obtained using the same ZPE correction procedure as for diamond.
Single-determinant DMC with a PBE0 trial wave function gives 4.683(3) eV, in excellent agreement with the experimental results, albeit relying on the cancellation of the atomic fixed-node error~\cite{Annaberdiyev2021May}. DFT performs reasonably well: PBE and HSE predict cohesive energies of 4.56~eV and 4.58~eV, corresponding to relative errors of 3-4\%. Among correlated wave-function methods, performance is mixed: MP2 (4.96~eV) is slightly worse than PBE, whereas CCSD (4.15~eV) underbinds substantially, consistent with the trend observed in diamond. {Previous AFQMC result based on Gaussian orbitals significantly underestimated the cohesive energy due to finite-size and finite-basis-set errors}~\cite{morales2020accelerate}. Prior plane-wave AFQMC gives decent agreement with experiment, although finite-size effects were addressed using the LDA correction~\cite {zhangQuantumMonteCarlo2003}. 

\begin{figure}
    \centering
    \includegraphics[width=0.9\linewidth]{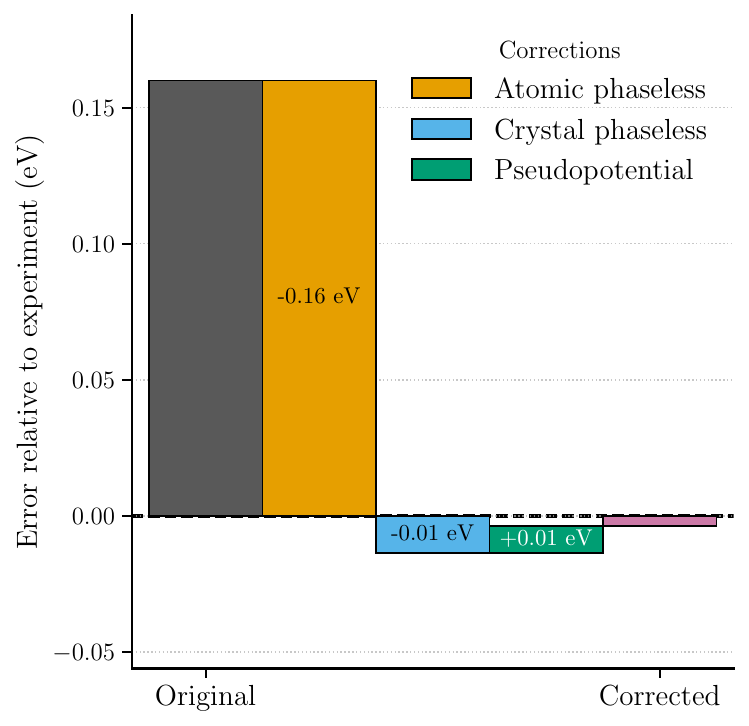}
    \caption{Errors in AFQMC from different sources, including atomic phaseless, crystalline phaseless and pseudopotential error in the cohesive energy of silicon. The atomic phaseless error was corrected using SHCI for TZ and QZ basis sets, and the results were extrapolated to the complete basis set limit. The crystalline phaseless error was corrected using CISD-AFQMC on a $2\times2\times2$ supercell in the DZ basis. The pseudopotential error was corrected using a $3\times3\times3$ supercell at the DZ level. The error bars for the corrections are not shown, as each is smaller than 0.01~eV, which would increase the final uncertainty by less than 0.01~eV.
    }
    \label{fig:erranalysis_Si}
\end{figure}

Interestingly, despite the structural similarity between the two systems, our AFQMC performs markedly worse for silicon. 
Our single-determinant AFQMC yields a cohesive energy of 4.89(2) eV with the GTH-HF-rev pseudopotential, which corresponds to a deviation of 0.16(2) eV compared to the more recent experimental value. This is significantly larger than the errors observed for diamond.
According to our error analysis in \cref{fig:erranalysis_Si}, the dominant source of error arises from the atomic phaseless bias. 
The pseudopotential error for silicon is considerably smaller than that for diamond.
Hence, there is no fortuitous cancellation of error in silicon.
This explains why AFQMC, without any corrections, yields accurate results for diamond but performs poorly for silicon. 
After correcting the atomic energies, we obtain a significantly improved cohesive energy for silicon, as shown in Fig.~\ref{fig:siliconerr}. 
{We note that this correction would lead to a much worse cohesive energy for the prior pw AFQMC result, suggesting that their LDA-based finite-size correction is ill-behaved for this example.}

In summary, single-determinant AFQMC performs reasonably well for covalent crystals with almost no multi-reference character. These analyses suggest that if all-electron calculations for all system sizes were performed (which we defer to future work) and the atomic energies were corrected using higher-level solvers, AFQMC could achieve quantitatively accurate cohesive energies for covalent solids even without explicit correction of the crystalline phaseless error.

\subsection{Metals}
Perturbative methods break down for metals because they involve energy denominators that typically vanish in metallic systems, ultimately leading to divergent energies. 
These methods include even the gold-standard coupled-cluster method, CCSD(T). This limitation of CCSD(T) spurred new approximate triples corrections in recent years~\cite{Neufeld2023Oct, masiosAverting2023}.
As a fully non-perturbative approach, AFQMC can, in principle, handle electron correlation in metallic systems without any extra care. 
Indeed, Lee {\it et al.} demonstrated that AFQMC yields accurate results for the uniform electron gas at high density ($r_s\le2.0$), although its accuracy degrades in the low-density regime~\cite{leeAuxiliaryField2019}.

{There are additional complications that arise for metallic systems. First, we have non-uniform electronic occupation over different $\wv$-points in the HF trial wavefunction. To efficiently perform dense matrix operations, we pad the coefficient matrix with zeros at $\wv$-points with occupation below the maximum. 
This is also related to the fact that real solids cannot naturally reach the ``magic number'' occupation like in the uniform electron gas model. The correlation energy change between different, energetically close occupations was found to be negligible, similar to what was observed in DMC~\cite{maezonoQuantum2003}
Next, we found that the equilibration time of AFQMC was much longer than that of semiconductors, as expected given its near-zero gap. To handle this, we employed a much larger time step for equilibration, after which we reset walker weights and started a production run with a smaller time step. More details are available in \cref{app:metals} with relevant numerical results.}

\begin{figure}
    \centering
    \includegraphics[width=1.0\linewidth]{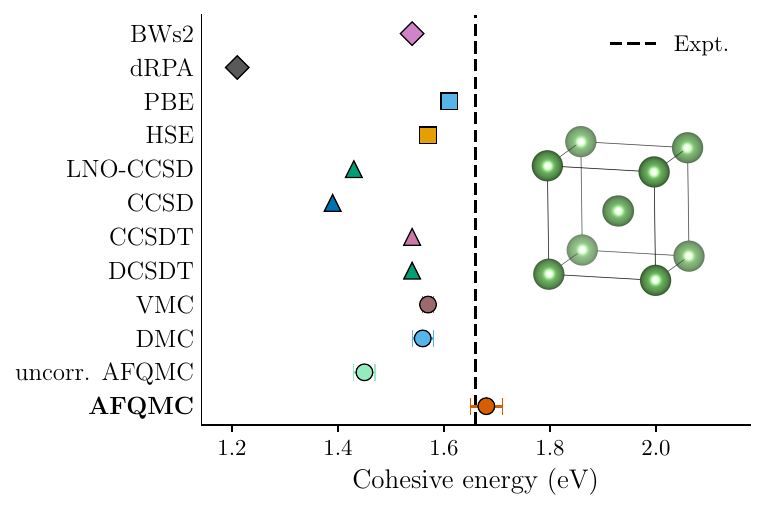}
    \caption{Error in the cohesive energy of BCC lithium predicted by different methods relative to experiment~\cite{chase1982janaf, Schimka2011Jan}. 
The BWs2 value is from Ref.~\cite{chenRegularized2025}; The dRPA value is from Ref.~\cite{Ye2024Oct}; PBE and HSE values are taken from Ref.~\cite{Schimka2011Jan}; The CCSD value is from Ref.~\cite{Neufeld2022Aug}; The CCSDT and DCSDT values are from Ref.~\cite{Neufeld2023Oct}, and the LNO-CCSD data is from Ref.~\cite{Ye2024Oct}; VMC is from Ref.~\cite{Yao1996Sep}, and the DMC result is from Ref.~\cite{Rasch2015Jul}. AFQMC result is from this work, which is corrected for atomic phaseless, crystalline phaseless and pseudopotential errors as described in the main text. The uncorrected AFQMC value is denoted as uncorr. AFQMC in the plot and is also given in Appendix \ref{cohdata}.
}
    \label{fig:lithiumerr}
\end{figure}

\subsubsection{BCC Lithium}
We report the cohesive energy of BCC lithium with an experimental lattice constant of 3.453 \AA~\cite{Schimka2011Jan} using AFQMC, a metal with a valence electron density corresponding to $r_s \approx $ 3.2. 
The experimental value of cohesive energy is 1.66 eV (ZPE corrected).
As can be seen in \cref{fig:lithiumerr}, 
DFT performs reasonably well, with PBE and HSE predicting 1.61 eV and 1.57 eV, respectively, both very close to the experimental value. {This is not surprising because PBE is a non-empirical GGA functional constrained to recover the uniform-electron-gas limit and related exact conditions. 
Since Li is a nearly ``uniform'' metal with relatively weak density inhomogeneity in the valence region, jellium-based constraints tend to be more appropriate, which explains why PBE often performs well for alkali metals.}

On the other hand, coupled-cluster methods span from 1.39 eV (CCSD) to 1.54 eV (coupled cluster with singles, doubles, and triples, CCSDT, and distinguishable cluster with singles, doubles, and triples, DCSDT), depending on the excitation level. 
In contrast, the direct random-phase approximation (dRPA) yields 1.21 eV, which is substantially below the experimental value. The size-consistent Brillouin-Wigner (BWs2) approach gives a good prediction of 1.54 eV. DMC and variational Monte Carlo (VMC) report 1.56(2) eV and 1.57(1) eV, respectively, both within 0.1 eV of the experimental value. 

Our single-reference AFQMC calculation gives a cohesive energy of 1.45(2) eV after direct TDL and CBS extrapolation. This corresponds to a deviation of $\sim$0.2 eV, somewhat larger than what we observed for the previous covalent crystals but still comparable to other state-of-the-art many-body methods.
Single-reference AFQMC clearly improves upon HF, dRPA, and CCSD but exhibits a larger deviation from experiment than DFT, DMC, and higher-order CC methods. 

We performed an error analysis of BCC lithium (see \cref{detailerroranalysis} for details). Because the Li atom has a ground-state total angular momentum $S$, AFQMC attains chemical accuracy at the atomic level~\cite{leeTwentyYearsAuxiliaryField2022}. 
A crystalline phaseless error is observed when comparing our results to CISD-AFQMC for the $2\times2\times2$ supercell at the DZ level, where we found a correction of approximately 1.1~m$E_h$ (0.03 eV) to the cohesive energy attributable to the crystalline phaseless bias.
Next, we found that the pseudopotential error, corrected at the $3\times 3 \times 3$ DZ level, is 0.2033(5)~eV. 
Applying these corrections yields a final AFQMC cohesive energy of 1.68(3)~eV, which is in excellent agreement with the experimental value. 
Interestingly, even the most costly methods considered here, such as CCSDT, do not yield quantitatively accurate cohesive energies for lithium, possibly also due to pseudopotential error in addition to residual finite-basis and size effects.

\subsubsection{FCC Aluminium}
Another metallic system studied here is face-centered cubic (FCC) aluminum, for which some benchmark data are available for comparison. 
Compared to lithium, the aluminum crystal exhibits stronger covalent bonding between atoms and is therefore less metallic in character. 
We model FCC aluminum using a conventional supercell containing four Al atoms with an experimental lattice constant 4.018 \AA~\cite{Schimka2011Jan}. However, due to the well-known shell effects~\cite{linTwistaveraged2001,holzmannTheory2016}, even at the HF level, it is challenging to converge the total energies to the TDL. 
To address this, we employ twist averaging~\cite{linTwistaveraged2001,drummondFinitesize2008,mihmShortcut2021} to accelerate convergence at the HF level (see Appendix~\ref{app:extraptdl} for details). 

\begin{figure}
    \centering
    \includegraphics[width=1.0\linewidth]{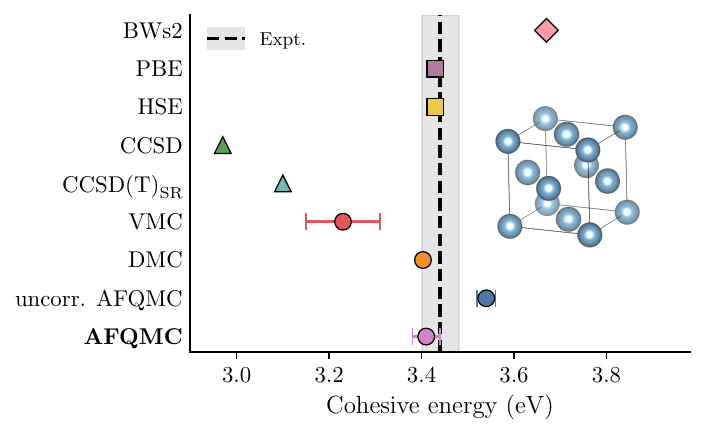}
    \caption{Error in the cohesive energy of FCC aluminum predicted by different methods relative to experiment~\cite{chase1982janaf, Schimka2011Jan}. The BWs2 value is from Ref.~\cite{chenRegularized2025}; PBE and HSE values are taken from Ref.~\cite{Schimka2011Jan}; CCSD and CCSD(T)$_\mathrm{SR}$ values are from Ref.~\cite{Neufeld2022Aug}; VMC is from Ref.~\cite{gaudoinInitio2002}, and the DMC result is from Ref.~\cite{hoodDiffusion2012}. The AFQMC result reported here is corrected for atomic phaseless error, as described in the main text. The uncorrected AFQMC value is denoted as uncorr. AFQMC in the plot and is also given in Appendix \ref{cohdata}.}
    \label{fig:aluminumerr}
\end{figure}

Subsequent AFQMC calculations based on the HF solution without twists exhibit a smooth extrapolation of the correlation energy following the $1/N_k$ behavior. With size extrapolation, the resulting cohesive energy is 3.54(2)~eV, compared to the experimental value of 3.43~eV, which corresponds to an error of approximately 0.11~eV (see Fig.~\ref{fig:aluminumerr}). Both CC methods significantly underestimate the cohesive energy, whereas PBE and HSE are nearly exact.  

Our AFQMC error analysis reveals that the dominant source of error is the atomic phaseless bias, since the ground state of the Al atom has total orbital angular momentum $L=P$~\cite{kaufmanWavelengths1991}. After correcting for the atomic phaseless error, the cohesive energy improves to 3.41(2)~eV, which is within 0.02~eV of the experimental value. 
Unfortunately, the CCSD calculation for the $2\times2\times2$ supercell could not converge, preventing a direct comparison between CISD-AFQMC and our AFQMC results. 

In summary, with appropriate handling of pseudopotential errors and an atomic-phaseless bias, single-determinant AFQMC can achieve near-exact accuracy in calculating the cohesive energy of prototypical metals.

\subsection{Transition metal oxides}
Transition-metal oxides (TMOs) are paradigmatic correlated materials in which partially filled $d$ shells give rise to competing energy scales and nontrivial magnetic order~\cite{imadaMetalinsulator1998,tokuraOrbital2000,heldElectronic2007}. As a consequence, DFT often struggles to predict their properties~\cite{kotliarElectronic2006,anisimovFirstprinciples1997,jiangFirstprinciples2015,wangLocal2019} and additional empirical corrections such as $+U$ are necessary for qualitatively correct answers~\cite{anisimovBand1991,himmetogluHubbardcorrected2014}. AFQMC, by contrast, has demonstrated superb performance in moderately to strongly correlated molecular systems~\cite{al-saidiAuxiliaryfield2006,mottaInitioComputationsMolecular2018,sheePotentially2023}, motivating its application to TMOs. 

In this work, we focus on the Heisenberg superexchange parameter $J$, which has been an important quantity for understanding magnetic order in correlated materials. We evaluate $J$ in two prototypical systems: rock-salt NiO and infinite-layer \ce{CaCuO2} (CCO), with the latter being an infinite-layer parent cuprate used as a minimal model for \ce{CuO2}-plane physics~\cite{siegristParent1988}.

We calculate the Heisenberg exchange coupling constant using an explicit broken-symmetry approach. In other words, we evaluate the energies of various spin-symmetry broken phases from which the Heisenberg exchange parameters are obtained by fitting to a Heisenberg model,
\begin{equation}
    \ham = -\sum_{i, j} J_{ij}\hat{\mathbf{S}}_i\cdot \hat{\mathbf{S}}_j.
\end{equation}

\subsubsection{NiO}

\begin{figure}
    \centering
    \includegraphics[width=0.5\linewidth]{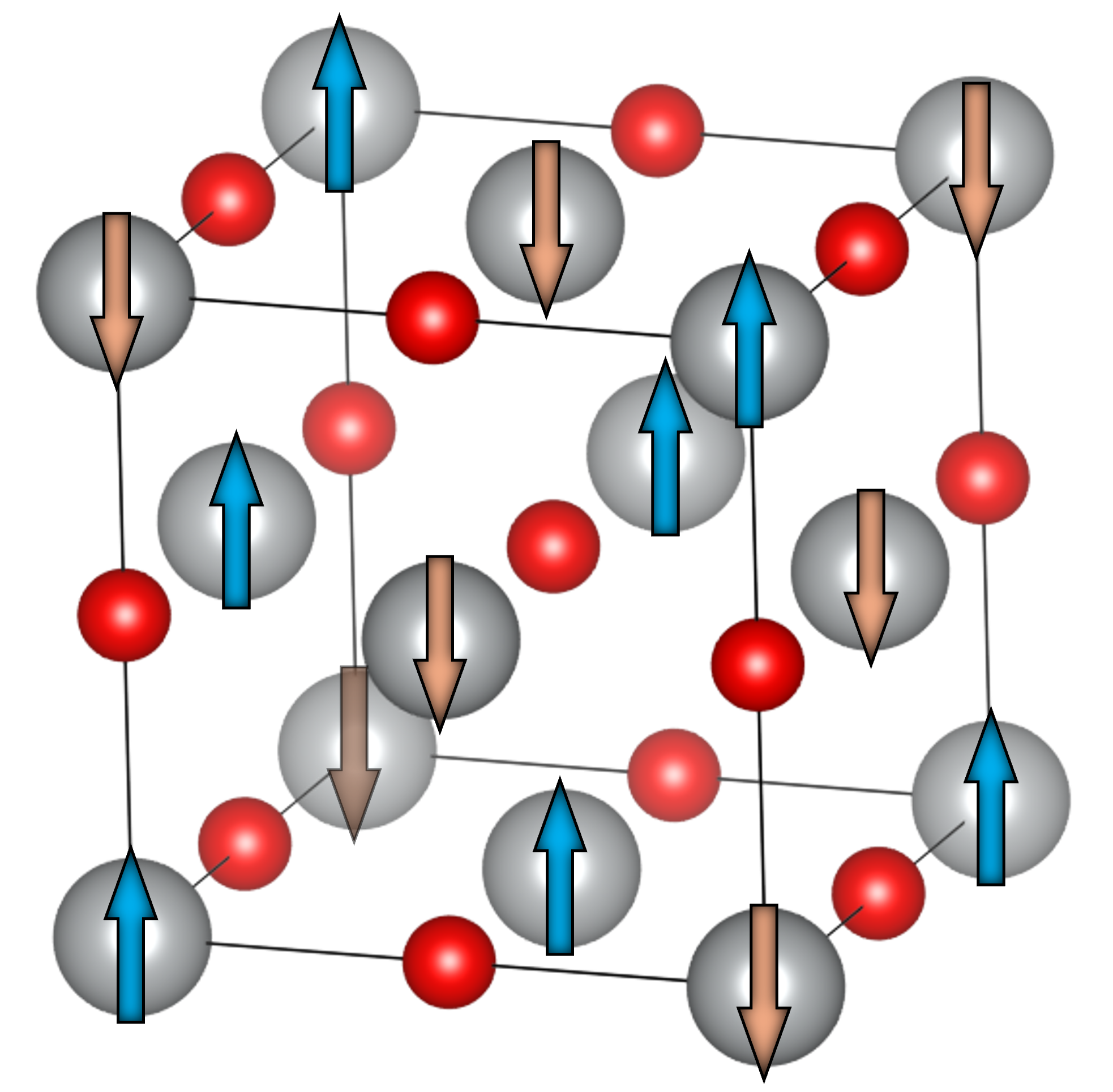}
    \caption{Illustration of the spin alignment in the AFMII phase of NiO.}
    \label{fig:niocrys}
\end{figure}

For NiO, a nearest-neighbor (NN) Heisenberg Hamiltonian is inadequate. Inelastic neutron scattering shows that the dominant coupling is the antiferromagnetic next-nearest-neighbor (NNN) exchange, $J_2$, and the NN coupling, $J_1$, is much smaller.~\cite{hutchingsMeasurement1972} 
In fact, the FCC Ni sublattice is frustrated with respect to NN exchange interactions, so the magnetic ground state is determined by the NNN exchange interactions~\cite {chatterjiAntiferromagnetic2009}. 
Therefore, we employ a Heisenberg model including both NN and NNN interactions to extract the NN exchange parameter $J_1$ and the NNN exchange parameter $J_2$. The exchange parameters can be calculated as~\cite{dep.r.moreiraEffect2002}
\begin{equation}
    J_1 = \frac{1}{8} (E_{\mathrm{AFMI}} - E_{\mathrm{FM}}),
\end{equation}
\begin{equation}
    J_2 = \frac{1}{6} (E_{\mathrm{AFMII}} - E_{\mathrm{FM}}) - J_1,
\end{equation}
where AFMII is the ground state phase in which the spins are ferromagnetically aligned within (111) planes and alternate between successive (111) layers, as shown in Fig. \ref{fig:niocrys}, whereas for AFMI the spins are ferromagnetically aligned within (001) planes and alternate between adjacent (001) layers. FM is the ferromagnetic phase in which all spins are parallel throughout the crystal.

We obtained Heisenberg exchange parameters from total-energy differences between distinct magnetic phases, computed independently using AFQMC. Because the energy differences are small (i.e., at the milliHartree level), it might be helpful to use the correlated sampling approach~\cite{sheeChemical2017,chenAlgorithm2023,otisScalable2025}. 
Unfortunately, correlated sampling is not readily applicable in this context, especially because population control decisions differ significantly across magnetic phases. Computational details are given in \cref{app:NiO}.

The calculated Heisenberg coupling constant is listed in Table \ref{tab:nio_coupling}. 
\begin{table}[htbp]
    \centering
    \caption{Exchange couplings $J_1$ and $J_2$ calculated using different methods. Fock-35 means that the exchange functional consists of $35\%$ Fock exchange and $65\%$ Dirac-Slater exchange. 
    }
    \begin{tabular}{cccc}
        \toprule
        Method & $J_1$ (meV) & $J_2$ (meV) & Reference\\
        \midrule
        HF & 0.8 & $-4.6$& \multirow{6}{*}{\cite{dep.r.moreiraEffect2002}}\\
        LDA & 11.9 & $-71.3$ & \\
        B3LYP & 2.4 & $-26.7$ & \\
        Fock-35 & 1.9 & $-19.7$ & \\
        CASSCF & 0.5 & $-5.0$ & \\
        CASPT2 & 1.2 & $-16.7$& \\
        scGW & 1.63 & $-13.61$ & \cite{pokhilkoBrokensymmetry2022}\\
        QSGW (+approx.)& $-0.8$ & $-14.7$ & \cite{kotaniSpin2008}\\
        DMET (CCSD) & / & $-14.05$ & \cite{yangInitio2026}\\
        AFQMC (downfolded) & / & $-18.0(5)$ & \cite{maQuantum2015,notenio}\\
        HF & 0.7 & $-$4.6 & This work \\ 
        AFQMC &$-0.4(1.7)$& $-19(2)$& This work\\
        \midrule
        \multirow{2}{*}{Expt.} & 1.4& $-19.0(3)$& \cite{hutchingsMeasurement1972}\\
         & $-1.4$& $-17.3$& \cite{shankerAnalysis1973a}\\
        \bottomrule
    \end{tabular}
    \label{tab:nio_coupling}
\end{table}
{The DFT results depend strongly on the fraction of exact exchange, and by tuning this parameter, one can get a broad range of values. The CASSCF and CASPT2 calculations were performed on cluster models of \ce{Ni2O10} or \ce{Ni2O11}. CASSCF yields values close to UHF, whereas including second-order perturbative corrections in CASPT2 leads to a substantial improvement over CASSCF. Self-consistent GW (scGW) yields results similar to those of quasiparticle self-consistent GW (QSGW), which both underestimate the magnitude of $J_2$. DMET with a CCSD impurity solver yields exchange couplings comparable to those obtained with GW. }

For our AFQMC, we determine the exchange couplings by extrapolating correlation energies to the TDL at the DZ level using $3\times3\times3$ and $4\times4\times4$ $\wv$-meshes (108 and 256 atoms, respectively). With the $3\times3\times3$ $\wv$-mesh, we obtained $J_1=-0.5\pm0.6\ \mathrm{meV}$ and $J_2=-30(1)\ \mathrm{meV}$. Because $|J_1|$ lies below our statistical error bar, we can only conclude that $J_1$ is small in magnitude, consistent with expectations. The apparent overestimation of $|J_2|$ at $3\times3\times3$ is resolved by proper TDL extrapolation, yielding $J_2=-19(2)\ \mathrm{meV}$. 
Hence, a size correction for exchange coupling is important.

\begin{figure}
    \centering
    \includegraphics[width=0.8\linewidth]{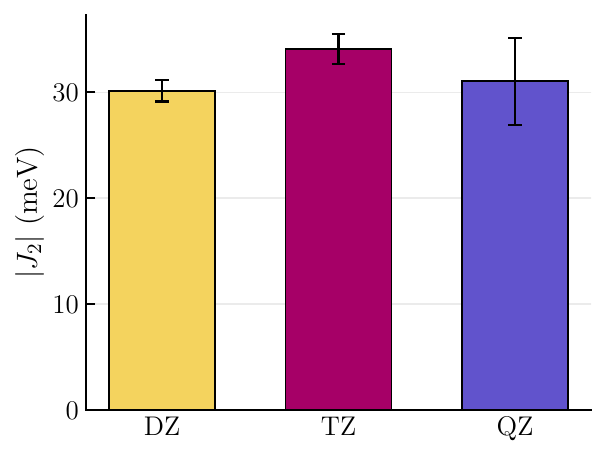}
    \caption{NiO $|J_2|$ values calculated on $3\times 3\times3$ supercell using different basis set sizes. Note that the difference we see between $J_2$ here and that in \cref{tab:nio_coupling} is due to the finite-size effect of the $3\times3\times3$ supercell.}
    \label{fig:J2bar}
\end{figure}

To assess the basis-set effects, we repeat the calculation with triple- and quadruple-zeta bases at $3\times 3 \times 3$ $\wv$- mesh and only found small changes in the coupling parameters, as shown in Fig. \ref{fig:J2bar}. We also performed an all-electron check at $3\times3\times3$ with a DZ basis set, and the resulting pseudopotential error is small (see Appendix \ref{detailerroranalysis} for details). 

\begin{figure}
    \centering
    \includegraphics[width=0.95\linewidth]{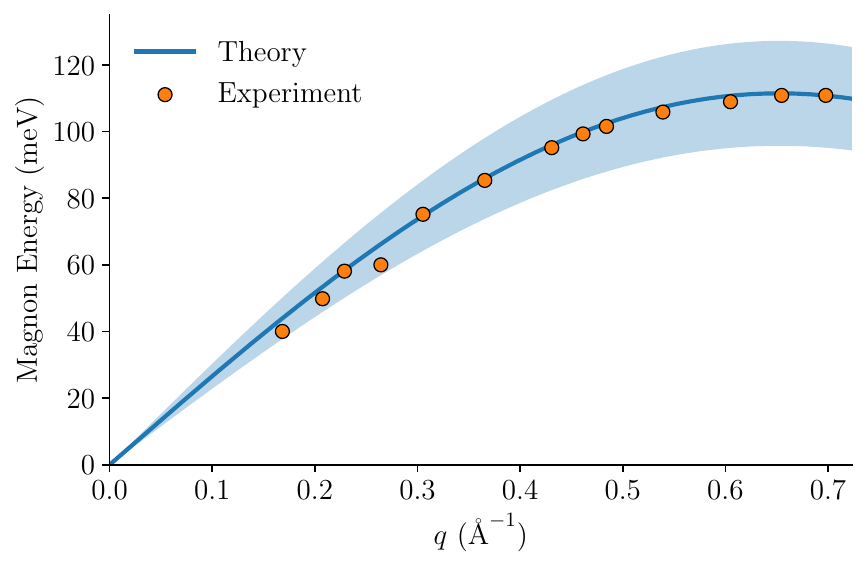}
    \caption{Magnon spectrum of NiO along the [111] direction computed using a NN+NNN Heisenberg model with exchange parameters extracted from AFQMC. The blue line indicates the mean value, and the shaded region denotes $\pm 1\sigma$. Experimental data are taken from Ref.~\cite{hutchingsMeasurement1972}.
    }
    \label{fig:magnonnio}
\end{figure}

Using these parameters, the NNN Heisenberg model reproduces the NiO magnon spectrum as measured by inelastic neutron scattering within statistical uncertainty, as shown in Fig.~\ref{fig:magnonnio}. We note that Ma et al. reported AFQMC results with downfolded Hamiltonians~\cite{maQuantum2015} using planewaves and DFT-based finite-size corrections~\cite{kweeFiniteSize2008,maFinitesize2011}. In contrast, our approach avoids such DFT-based finite-size corrections and also allows for the assessment of pseudopotential errors relative to all-electron calculations.

Lastly, we calculate the magnetic moment on the Ni atoms by calculating the Mulliken spin population defined as 
\begin{equation}
    m_{\mathrm{Ni}_A} = \sum_{\mu \in \mathrm{Ni}_A}\sum_{\wv, \nu} (P^\wv_\alpha - P^\wv_\beta)_{\mu\nu} S_{\nu\mu}^\wv
\end{equation}
Since this can be converted to the expectation value of a one-body operator defined as
\begin{equation}
    \hat{O} = \sum_{\mu \in \mathrm{Ni}_A}\sum_{\wv, \nu}S^\wv_{\nu\mu}(a^\dagger_{\mu\wv\alpha} a_{\nu\wv\alpha} - a^\dagger_{\mu\wv\beta}a_{\nu\wv\beta})
\end{equation}
we evaluate this expectation value by calculating the response estimator~\cite{mahajanResponse2023,ledermullerLocal2013,pinskiCommunication2018,zhangPerformant2024}
\begin{equation}
    \av{\hat{O}} \approx \frac{\d}{\d \lambda} \av{\hat{H} + \lambda \hat{O}} = \frac{\d E_\lambda}{\d \lambda} \approx \frac{E_\vep - E_0}{\vep}
\end{equation}
where $\vep$ is a small real constant, and we take $\vep = 10^{-5}$ in our study. The difference between $E_\varepsilon$ and $E_0$ is obtained with correlated sampling. {Correlated sampling is applicable here because $E_{\varepsilon}$ and $E_{0}$ correspond to the same magnetic phase and differ only by a small perturbation.} For the AFMII ground state, as shown in \cref{tab:nio_magmom}, we obtain a local magnetic moment of $m = 1.69(3)\ \mathrm{a.u.}$, in very good agreement with experiment and other many-body approaches such as dynamical mean-field theory (DMFT) and CCSD. It is also known that HF tends to overestimate the magnetic moment because it overemphasizes an ionic contribution that stabilizes a larger local moment~\cite{chenNature2012}. This bias is substantially reduced once electron correlation is included, as evidenced by the markedly smaller magnetic moment obtained with AFQMC.

\begin{table}[htbp]
    \centering
    \caption{Local magnetic moment on Ni in the AFMII ground state of NiO computed using various methods, together with experimental values. See Appendix~\ref{app:NiO} for details.
    }
    \begin{tabular}{ccc}
        \toprule
        Method & $m_{\mathrm{Ni}}$ ($\mu_{\mathrm{B}}$) & Reference\\
        \midrule
        PBE & 1.34 & \multirow{2}{*}{\cite{gaoElectronic2020}}\\
        CCSD & 1.72 & \\
        GW & 1.83 & \cite{massiddaQuasiparticle1997}\\
        LDA + $U$ & 1.70 & \cite{anisimovDensityfunctional1993}\\
        LDA+DMFT & 1.70 & \cite{renLDA2006}\\
        GW+DMFT& 1.69 & \cite{zhuInitio2021}\\
        HF & 1.78 & This work\\
        AFQMC & 1.69(3) & This work\\
        \midrule
        \multirow{3}{*}{Expt.} & 1.64 & \cite{alperin1962magnetic}\\
         & 1.77 & \cite{fenderCovalency1968}\\
         & 1.90 & \cite{cheethamMagnetic1983}\\
        \bottomrule
    \end{tabular}
    \label{tab:nio_magmom}
\end{table}

\subsubsection{\ce{CaCuO2}}
\begin{figure}
    \centering
    \includegraphics[width=0.7\linewidth]{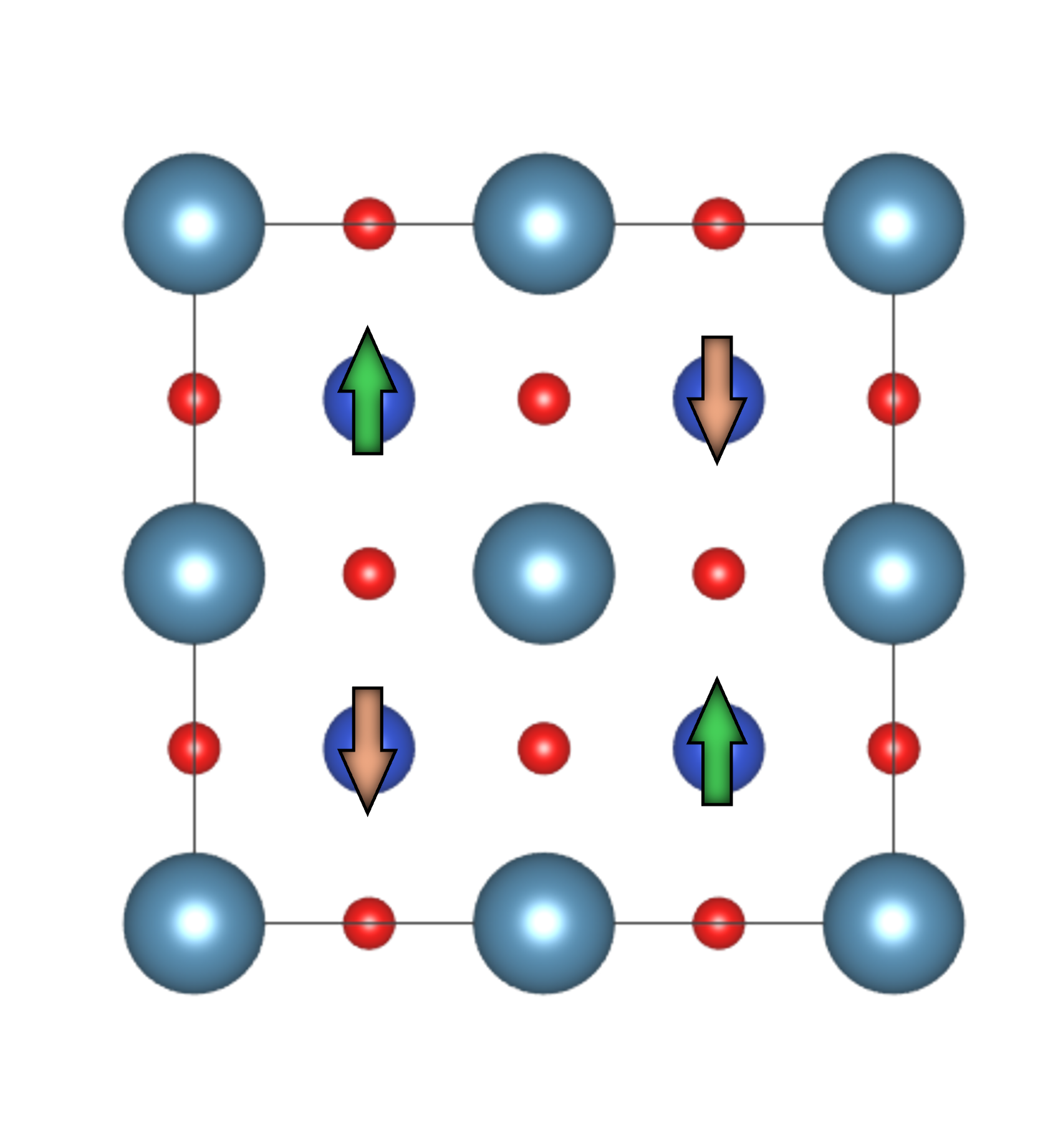}
    \caption{Illustration of the spin alignment in the AFM phase of \ce{CaCuO2}, the structure is shown as a projection along the $z$ axis. }
    \label{fig:magcco}
\end{figure}

As a cuprate parent compound with a clean, simple crystal structure, CCO and related materials (such as \ce{Ca2CuO3}) have attracted substantial interest in computational studies~\cite{foyevtsovaInitio2014,wagnerEffect2014,cuiSystematic2022}.
CCO has a magnetic structure different from that of NiO: it has a layered structure with most of the strong correlation physics contained in the \ce{[CuO2]^{2-}} plane. 
The magnetic structure of CCO, as shown in \cref{fig:magcco}, can be modeled by a 2D NN Heisenberg or a 2D one-band Hubbard model. 
Therein, the NN exchange coupling $J$ is calculated as
\begin{equation}
    J = \frac{1}{4}(E_{\mathrm{AFM}} - E_{\mathrm{FM}}),
\end{equation}
for both models.
Based on the observation that the finite-size effect is pronounced in NiO, we also perform extrapolation to the TDL. It is a more stringent test than NiO, since electron correlation effects are much more complex in it. Due to its quasi-2D nature, we used an anisotropic $\wv$-mesh as suggested in Ref.~\cite{cuiSystematic2022}. We provide more computational details in \cref{app:cco}. 

\begin{table}[htbp]
    \centering
    \caption{NN exchange coupling $J$ (meV) for \ce{CaCuO2} computed using various methods.  
    }
    \begin{tabular}{ccc}
    \toprule
    Method & $-J$ (meV) & Reference\\
    \midrule
    HF & 38.0 & \multirow{4}{*}{\cite{cuiSystematic2022}}\\
    PBE$+ U$ & 168.9 & \\
    PBE0 & 213.9 & \\
    DMET(CCSD) & 155.4 & \\
    DMET(CCSD) & 165.44 & \cite{yangInitio2026} \\
    DMC & 140(20)& \cite{wagnerEffect2014}\\
    HF & 38.1 & This work\\
    AFQMC & 190(15)& This work\\
    \hline
    \multirow{3}{*}{Expt. (Heisenberg)} & 142 & \cite{kanPreparation2004}\\
     & 157 & \cite{minolaMagnetic2012}\\
     & 158 & \cite{pengInfluence2017}\\\hline
     \multirow{2}{*}{Expt. (Hubbard)} & 182 & \cite{pengInfluence2017}\\
     & 172(7) & \cite{martinelliFractional2022}\\
    \bottomrule
    \end{tabular}
    \label{tab:ccoj}
\end{table}

In \cref{tab:ccoj}, we present the exchange coupling of various methods. DFT methods such as PBE$+U$ and PBE0 significantly improve upon Hartree-Fock results, but they either underestimate or overestimate $|J|$ relative to the one-band Hubbard model parameter. Similarly, both DMET(CCSD) and DMC also slightly underestimate the exchange coupling.
AFQMC yields $|J| = 190(15)$ meV, which is higher than the NN-Heisenberg experimental estimates (142-158 meV), though within 2$\sigma$. In contrast, it lies comfortably within the range of exchange parameters extracted from the one-band Hubbard model, which provides a more correct representation of the measured spin-wave spectrum in \ce{CaCuO2} as shown in Ref.~\cite{pengInfluence2017}. 

Because the exchange parameter $J$ is inferred from experiment by fitting to a chosen low-energy model, its numerical value is model-dependent. In the case of the single-band Hubbard model, it naturally incorporates effects of the interactions absent in the NN Heisenberg model. 
The closer agreement with the exchange parameter from the one-band Hubbard model thus suggests that AFQMC captures sophisticated electron-correlation effects that include longer-range and cyclic exchange processes in addition to the simple NN exchange.
One can, in principle, obtain the cyclic exchange coupling, $J_c$, to calculate the full magnon spectrum for CCO using AFQMC.
While we expect this to be relatively straightforward, the required error bar in each energy calculation
would be smaller than what we needed for $J$, and hence, it will demand substantially higher computational cost. Therefore, we leave a more exhaustive investigation of the magnetic properties of the normal state of high-$T_c$ materials for future work.

\section{Conclusions}
In this work, we developed a scalable AFQMC algorithm for solid-state systems with Gaussian basis sets. By combining the low-rank factorization of electron-repulsion integrals with $\wv$-point symmetry, we removed the dominant memory bottleneck in conventional Gaussian-based AFQMC simulations. 
Further, we introduced new contraction schemes for both the propagation and local-energy evaluations to significantly accelerate the computation. 
The method is implemented on GPUs, enabling large-scale simulations of realistic solids. 
As a result, we can obtain energies for simple solids at the TDL and CBS limits without relying on empirical corrections or finite-size corrections imported from lower-level methods.

We validated the approach on semiconductors (diamond, silicon), metals (BCC lithium, FCC aluminum), and correlated transition metal oxides (NiO and CCO). Through a detailed error analysis, we identified the main sources of error in the semiconductor and metallic benchmarks. By correcting those errors, we showed that AFQMC reproduces experimental cohesive energies within statistical uncertainty, demonstrating greater accuracy than state-of-the-art many-body methods. 
For NiO, we found that finite-size effects are critical for obtaining the correct Heisenberg exchange couplings, and the computed spin moment is consistent with experiment. {For CCO, our exchange coupling is close to the estimates based on a one-band Hubbard model, suggesting that AFQMC captures more sophisticated electron correlation effects beyond a strictly NN spin model.}

Eliminating the primary bottlenecks in solid-state AFQMC with a Gaussian basis set, our work establishes a scalable path toward accurate, benchmark-quality simulations of insulating, metallic, and strongly correlated quantum materials. In this context, AFQMC's favorable computational scaling and memory footprint make it a more practical alternative to CC approaches, and our benchmarks show that it consistently outperforms CCSD in accuracy. At the same time, because AFQMC works in an orbital representation rather than real space, it is naturally compatible with all-electron formulations, which then provides a clear route to completely eliminate pseudopotential errors that have plagued the performance of DMC.
Furthermore, it naturally provides a non-perturbative way to incorporate spin-orbit coupling via the exact two-component relativistic framework~\cite{liu2009exact}.
While we did not utilize any local correlation or embedding methods, we envision that the combination of AFQMC and these other local techniques will enable
many-body simulations at a significantly large scale with high accuracy.
{Hence, our findings suggest that AFQMC can be applied to even more complex systems such as high-$T_c$ materials and other exotic strongly correlated systems, with all material-specific details.}

Another research direction we will pursue is the investigation of finite-temperature AFQMC that has only been utilized for lattice models via constrained-path approximations~\cite{zhangFiniteTemperature1999,heFinitetemperature2019,shenFinite2020} and
uniform electron gas model with the phaseless approximation~\cite{leePhaseless2021}.
Combined with our algorithmic development presented here, we expect AFQMC to provide alternative routes to modeling warm-dense matter~\cite{grazianiFrontiers2014,vorbergerRoadmap2025}.
Furthermore, using this algorithm, we can calculate imaginary-time dynamical correlation functions that can be used to obtain spectral functions of real materials via analytic continuation~\cite{silverMaximumentropy1990,gubernatisQuantum1991,schummSingleparticle2025}.
Work along these lines is in progress in our group.

\section*{Data availability}
Data used in this work are online at~\cite{data_github}.

\begin{acknowledgments}
This work was supported by Harvard University's
startup funds and the DOE Office of Fusion Energy Sciences
``Foundations for quantum simulation of warm dense matter'' project. This research used computational resources of the National Energy Research Scientific Computing Center (NERSC), a Department of Energy User Facility using NERSC award ERCAP0034413 and ERCAP0037131, the Oak Ridge Leadership Computing Facility at the Oak Ridge National Laboratory through the Director's Discretion Project CHP135 and the INCITE project MAT281, which is supported by the Office of Science of the U.S. Department of Energy under Contract No. DE-AC05-00OR22725, and the FAS Research Computing cluster supported by the Faculty of Arts and Sciences (FAS) Division of Science Research Computing Group at Harvard. 
The work of H.Q.D. and A.R. was in part supported by the Breakthrough Energy Foundation.
The work of T.J. was in part supported by the Gordon and Betty Moore Foundation Postdoctoral Fellowship.
The work of A.Z.N. was in part supported by the NSF GRFP Fellowship.
The work of A.R. was in part supported by the Harvard Quantum Initiative prize postdoctoral fellowship. 
P.J.R. acknowledges the support of the Arnold and Mabel Beckman Foundation (\url{http://dx.doi.org/10.13039/100000997}) through the Arnold O. Beckman Postdoctoral Fellowship.
We thank Andrew Baczewski, Tim Berkelbach, Garnet Chan, Matthew Foulkes, Paul Kent, Boris Kozinsky, Fionn Malone, David Reichman, Sandeep Sharma, and Lucas Wagner for discussion and encouragement. J. Z. thanks W. C. Witt for his assistance in utilizing INCITE computational resources.
\end{acknowledgments}
\bibliography{references}
\clearpage
\appendix
\onecolumngrid
\setcounter{figure}{0}
\renewcommand{\thefigure}{A\arabic{figure}}
\renewcommand{\theHfigure}{A.\arabic{figure}}
\setcounter{table}{0}
\renewcommand{\thetable}{A\arabic{table}}
\renewcommand{\theHtable}{A.\arabic{table}}
\setcounter{algorithm}{0}
\renewcommand{\thealgorithm}{A\arabic{algorithm}}
\providecommand{\theHalgorithm}{\thealgorithm}
\renewcommand{\theHalgorithm}{A.\arabic{algorithm}}

\section{Additional details of $\wv$-THC-AFQMC}\label{app:afqmc}
This section complements the main text by detailing additional theoretical aspects of $\wv$-THC-AFQMC. 
\subsection{Cost reduction due to $\mathbf k$-point symmetry}
We can reduce the computation cost by half using the permutation symmetry of the ERI, following the approach discussed in Refs.~\cite{mottaHamiltonian2019,Malone2020May}, i.e. 
\begin{equation}
    (p\wv, r\wv + \qv| q \wv' + \qv, s\wv') = (r\wv + \qv, p \wv | s\wv', q\wv' + \qv)^*,
\end{equation}
and it follows that $M^\qv_{PQ}=M^{-\qv*}_{PQ}$, which then leads to the following symmetry of the $\hat{L}_{\gamma, \qv}$ operator:
\begin{equation}
    \hat{L}_{\gamma, \qv} = \hat{L}^\dagger_{\gamma, -\qv}.
\end{equation}
Thus, the operators $\hat{L}^+_{\gamma, \qv}$ and $\hat{L}_{\gamma, \qv}^-$ satisfy 
\begin{equation}
\begin{aligned}
    \hat{L}^+_{\gamma, -\qv} &= \hat{L}^+_{\gamma, \qv},\\
    \hat{L}^-_{\gamma, -\qv} &= -\hat{L}^-_{\gamma, \qv}.
\end{aligned}
\end{equation}
So it's possible to partition the set of $\qv$ points into 3 disjoint subsets $\mathcal{S} \cup \mathcal{Q}_+ \cup \mathcal{Q}_-$, where $\mathcal{S} = \{\qv : \exists \mathbf{G}\text{ such that }\qv = -\qv + \mathbf{G}\}$ is the “self-invariant’’ subset, and $\mathcal{Q}_{+}$ contains exactly one representative from each remaining pair ${\qv,-\qv}$, while $\mathcal{Q}_{-} = \{-\qv : \qv\in \mathcal{Q}{+}\}$. By construction, $\mathcal{S}\cup\mathcal{Q}_{+}\cup\mathcal{Q}_{-}$ exhausts all $\qv$ points, and the three subsets are pairwise disjoint. From the symmetry analysis above, we know that the two-body Hamiltonian can be written as
\begin{equation}
    \ham_2' = -\frac{1}{2}\sum_{\gamma,\mathbf{q}\in \mathcal{S}} ((\hat{L}^{+}_{\gamma, \mathbf{q}})^2 + (\hat{L}^{-}_{\gamma, \mathbf{q}})^2) - \sum_{\gamma,\mathbf{q}\in \mathcal{Q}_+}((\hat{L}^{+}_{\gamma, \mathbf{q}})^2 + (\hat{L}^{-}_{\gamma, \mathbf{q}})^2)
\label{eq:h2p_symm}
\end{equation}
hence we can define the reduced index set of $\qv$ points to be $\{\tilde{\qv}\} = \mathcal{S} \cup \mathcal{Q}_+$, and 
\begin{equation}
    \hat{\mathcal{L}}_{\gamma, \tilde{\qv}} = \begin{cases}
  \hat{L}_{\gamma,\tilde{\qv}}, & \tilde{\qv}\in\mathcal{S},\\[6pt]
  \sqrt{2}\,\hat{L}_{\gamma,\tilde{\qv}}, & \tilde{\qv}\in\mathcal{Q}_{+}.
\end{cases}
\end{equation}

\subsection{Force bias, mean field shift and phaseless approximation}\label{app:mfshift}
To suppress fluctuations and reduce phaseless bias, we apply a mean-field shift by subtracting the trial-state average of $\hat{L}_{\gamma,\qv}$ from the original operator: $\hat{L}_{\gamma,\qv} \to \hat{L}_{\gamma,\qv} - \ovl{L}_{\gamma,\qv}$, where $\ovl{L}_{\gamma, \qv} = \av{\Psi_T|\hat{L}_{\gamma, \qv}|\Psi_T}/\braket{\Psi_T|\Psi_T}$. Under the THC form of the ERI, one can show that
\begin{equation}
    \ovl{L}_{\gamma,\qv} = \delta_{\qv, \mathbf{0}}\sum_{P\wv}\sum_{pr\sigma} \psi_p^\wv(\pos_P)^* \psi_r^{\wv }(\pos_P)R^{\mathbf{0}}_{P\gamma}G^{\Psi_T \sigma}_{p\wv, r\wv},
\end{equation}
where $G^{\Psi_T \sigma}_{p\wv, r\wv}$ is the one-body Green's function evaluated using the trial wavefunction
\begin{equation}
    G^{\Psi_T \sigma}_{p\wv, r\wv} = \frac{\braket{\Psi_T|a^\dagger_{p\wv} a_{r\wv}|\Psi_T}}{\braket{\Psi_T|\Psi_T}}.
\end{equation} 
Since $M^\qv_{PQ}=M^{-\qv*}_{PQ}$ implies $M^{\mathbf{0}}_{PQ}$ is real, it follows that $\ovl{L^\dagger}_{\gamma,\qv}=\ovl{L}_{\gamma,\qv}$, hence $\ovl{L^+}_{\gamma,\qv}=i\ovl{L}_{\gamma,\qv}$ and $\ovl{L^-}_{\gamma,\qv}=0$.
Further, the optimal force bias is subtracted from the random fields in the Hubbard-Stratonovich decomposition to minimize variance, which is evaluated as 
\begin{equation}
\begin{aligned}
        \overline{x}_{\gamma, \tilde{\qv}} &= -\sqrt{\Delta\tau} \frac{\braket{\Psi_T|\hat{\mathcal{L}}_{\gamma, \tilde{\qv}}|\Phi}}{\braket{\Psi_T|\Phi}} \\
    &= -\sqrt{\Delta \tau}\sum_{\wv P ir\sigma}\mathcal{R}^{\tilde{\qv}, \gamma}_P \psi_{i}^{\wv *}(\pos_P) \psi^{\wv + \tilde{\qv}}_r(\pos_P)\mathcal{G}^{\sigma}_{i\wv, r\wv+\tilde{\qv}}
\end{aligned}
\end{equation}
where $\mathcal{R}^{\tilde{\qv}, \gamma}_P$ is the Cholesky decomposition of $M_{PQ}^\qv$ on the reduced $\qv$-point set, and we introduced the half-rotated Green's function
\begin{equation}
    G^\sigma_{p\wv_1, q\wv_2} = \sum_i [\Psi^\sigma_T]^{\wv_1*}_{pi}\mathcal{G}^\sigma_{i\wv_1, q\wv_2}
\end{equation}
given the matrix representation $[\Psi^\sigma_T]^{\wv_1*}_{pi}$ of the state $\ket{\Psi_T}$ in the orthogonal Bloch basis. Similar formulas can be obtained for the force bias $\overline{y}_{\gamma, \qv}$ corresponding to the auxiliary fields $y_{\gamma, \qv}$.

In free-projection AFQMC, the weights of the walkers are updated using the importance function (with importance sampling)
\begin{equation}
    w_\alpha (\tau + \Delta \tau) = w_\alpha (\tau) I(\mathbf{x}_\alpha(\tau), \ovl{\mathbf{x}}_\alpha(\tau), \ket{\Phi_\alpha(\tau)}),
\end{equation}
where $\mathbf{x}_\alpha$ denotes the collection of the auxiliary fields for walker $\alpha$, $\ovl{\mathbf{x}}_\alpha(\tau)$ denotes the corresponding force bias, and the importance function $I(\mathbf{x}_\alpha(\tau), \ovl{\mathbf{x}}_\alpha(\tau), \ket{\Phi_\alpha(\tau)})$ is defined as
\begin{equation}
    I(\mathbf{x}_\alpha, \ovl{\mathbf{x}}_\alpha,\ket{\Phi_\alpha}) = \frac{\braket{\Psi_T|\hat{B}(\mathbf{x}_\alpha - \ovl{\mathbf{x}}_\alpha)|\Phi_\alpha}}{\braket{\Psi_T|\Phi_\alpha}}\times \exp\left(\mathbf{x}_\alpha \cdot \ovl{\mathbf{x}}_\alpha -\frac{1}{2}\ovl{\mathbf{x}}_\alpha\cdot \ovl{\mathbf{x}}_\alpha \right).
\end{equation}
In practice, the phase problem causes the signal-to-noise ratio to decay exponentially. We therefore project the importance function onto a real, nonnegative factor,
\begin{equation}
    I_\mathrm{ph}(\mathbf{x}_\alpha, \ovl{\mathbf{x}}_\alpha,\ket{\Phi_\alpha}) = ||I(\mathbf{x}_\alpha, \ovl{\mathbf{x}}_\alpha,\ket{\Phi_\alpha})|| \times \max \{0, \cos \arg I(\mathbf{x}_\alpha, \ovl{\mathbf{x}}_\alpha, \ket{\Phi_\alpha})\},
\end{equation}
which defines the phaseless approximation. 

\subsection{Propagation algorithm}
\label{propdetails}
As noted in the main text, pre-contracting the Bloch basis functions with the walker wavefunction reduces the cost from $\mathcal{O}(N_k^2 M^2 ( N_{\mathrm{ISDF}} + N_k n_{\mathrm{occ}}))$ to $\mathcal{O}(N_k^2 N_{\mathrm{ISDF}} n_{\mathrm{occ}} (N_k + M))$. Hence, this matrix-vector contraction is often preferable to explicitly forming \(\mathbf{V}_{\mathrm{HS}}\) and applying it repeatedly in the Taylor expansion of the two-body propagator for the cases where $n_{\mathrm{occ}}$ is small, and thus we obtain an $n_{\mathrm{occ}}n_{\mathrm{Taylor}}/M$ saving, where $n_{\mathrm{Taylor}}$ is the order of the Taylor expansion of the matrix exponential. Moreover, for large systems, the storage cost of $\mathbf{V}_\mathrm{HS}$ scales as $\mathcal{O}(N_{\mathrm{walker}}N_{k}^2N_{\mathrm{bsf}}^2)$, which is expensive and substantially reduces the memory available for other intermediates. Therefore, this algorithm is more memory-efficient, since it stores only slices of the intermediates $X$ and $Y$ based on available memory and avoids building the full $\mathbf{V}_\mathrm{HS}$ matrix. We provide the pseudocode in \cref{algorithm:vhs}.

\begin{algorithm}[H]
\caption{Algorithm for $\hat{V}_{\mathrm{HS}}$-walker contraction}\label{algorithm:vhs}
\begin{algorithmic}[1]
\State $\displaystyle X^{\wv'', \wv'}_{iP} \gets \sum_{r}\psi^{\wv''}_{r}(\pos_P)[\Phi]_{r\wv'', i\wv'} $ 
    \Comment{$\mathcal{O}\left(N_{k}^2N_{\mathrm{ISDF}}n_{\mathrm{occ}}M\right)$}
    \State $\displaystyle Y^{\wv, \wv'}_{iP} \gets \sum_{\wv''}[Rx]^{\wv'' - \wv}_P X^{\wv'', \wv'}_{iP}$ \Comment{$\mathcal{O}\left(N_{k}^3N_{\mathrm{ISDF}}n_{\mathrm{occ}}\right)$}
    \State $\displaystyle[\hat{V}_{\mathrm{HS}}\Phi]_{p\wv, i\wv'} = \sum_P \psi_p^{\wv*} (\pos_P) Y^{\wv, \wv'}_{iP}$
           \Comment{$\mathcal{O}\left(N_{k}^2N_{\mathrm{ISDF}}n_{\mathrm{occ}}M\right)$}
\end{algorithmic}
\end{algorithm}

\begin{figure}
    \begin{small}
        \begin{center}
            \includegraphics[width=0.55\textwidth]{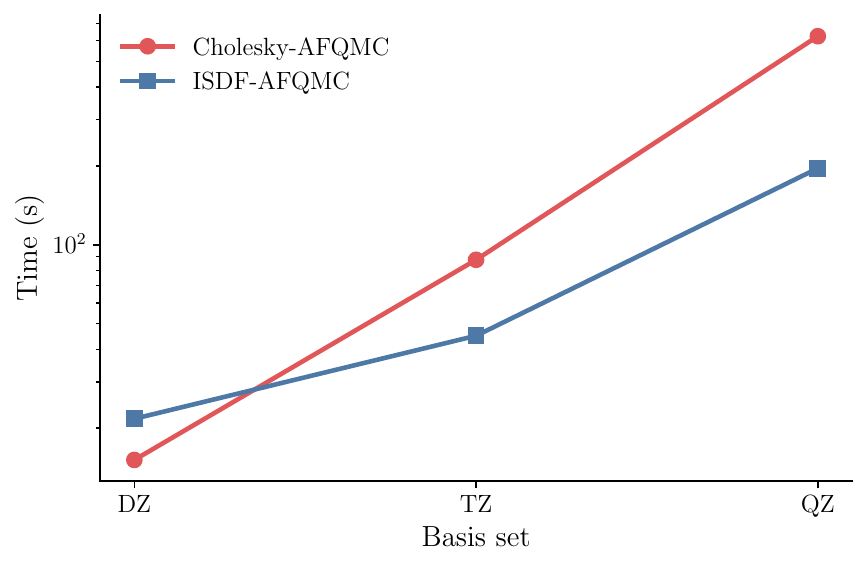}
        \end{center}
        \caption{Wall-time comparison between THC-AFQMC and Cholesky-AFQMC for diamond across multiple basis sets, sampled on a $3\times3\times3$ $\wv$-mesh. The test systems are small enough that local-energy evaluation is not rate-limiting. Timings were obtained on a single A100 GPU (80 GB), with AFQMC run with 10 walkers in a single process over 10 blocks.}
        \label{fig:isdfspeedup}
    \end{small}
\end{figure}

However, in the conventional Cholesky AFQMC formulation, no analogous low-cost contraction path exists. Consequently, for systems with few valence electrons, THC-AFQMC can be much faster than Cholesky AFQMC with the help of the algorithm above, as shown in Fig. \ref{fig:isdfspeedup}. This might be counterintuitive because, were \(\mathbf{V}_{\mathrm{HS}}\) constructed explicitly, THC would appear to incur additional contractions to assemble the Cholesky vectors. 

\subsection{Estimators}\label{app:estimators}
In practice, we sample the integral in Eq. \eqref{eq: HS} with importance sampling in the single-determinant space,
\begin{equation}
    \ket{\Phi_w} = \sum_\alpha^{N_\mathrm{w}}w_\alpha\frac{\ket{\Phi_\alpha}}{\braket{\Psi_T|\Phi_\alpha}}
\end{equation}
hence the energy estimates are obtained by the weighted sum of local energies
\begin{equation}
     E = \frac{\displaystyle\sum_\alpha^{N_\mathrm{w}}w_\alpha E_{L,\alpha}}{\displaystyle\sum_\alpha^{N_\mathrm{w}}w_\alpha}
\end{equation}
and the local energy $E_{L,\alpha}$ is given by
\begin{equation}
\begin{aligned}
    E_{L,\alpha} &= \frac{\braket{\Psi_T|\hat{H}|\Phi_\alpha}}{\braket{\Psi_T|\Phi_\alpha}}\\
    &= \sum_{pq\wv, \sigma} h^\wv_{pq} G^{\alpha,\sigma}_{p\wv, q\wv}\\
    &+ \sum_{\substack{pqrs\gamma,\wv\wv'\qv\\\sigma,\sigma'}}L_{p\wv , r\wv+ \mathbf{q}}^{\gamma, \mathbf{q}}L_{s\wv',q\wv' + \mathbf{q}}^{\gamma, \mathbf{q}*}(G^{\alpha,\sigma}_{p\wv, r\wv+\qv }G^{\alpha,\sigma'}_{q\wv' + \qv,s\wv'} \\
    &- \delta_{\sigma\sigma'}G^{\alpha,\sigma}_{p\wv, s\wv' }G^{\alpha,\sigma'}_{q\wv' + \qv,r\wv+\qv}),
\end{aligned}
\end{equation}
where 
\begin{equation}
    G^{\alpha,\sigma}_{p\wv, q\wv} = \frac{\braket{\Psi_T|\cre_{p\wv,\sigma}a_{q\wv,\sigma}|\Phi_\alpha}}{\braket{\Psi_T|\Phi_\alpha}}
\end{equation}
is the mixed Green's function. The evaluation of the local energy formally scales as $\mathcal{O}(N_k^3n^4)$, exhibiting a $\mathcal{O}(N_k)$ saving in the cost due to lattice translational symmetry. 

\subsection{Local energy algorithm using THC}
Using the half-rotated Green’s function yields an $n_{\mathrm{occ}}/M$ speedup in evaluating the local energy, which is given by
\begin{equation}
    \begin{aligned}
    E_{L} &= \sum_{ip\wv, \sigma} \mathcal{T}^{\sigma\wv}_{ip} \mathcal{G}^\sigma_{i\wv, p\wv}\\
    &+ \sum_{\substack{ijpq,\wv\wv'\qv\\\sigma,\sigma'}}\sum_{PQ}\psi^{\wv*}_{i\sigma}(\pos_P)\psi^{\wv+\qv}_{p}(\pos_P)M_{PQ}^\qv \psi_{j\sigma'}^{\wv'+\qv*}(\pos_Q)\psi_{q}^{\wv'}(\pos_Q)\\
     &(\mathcal{G}^\sigma_{i\wv, p\wv+\qv }\mathcal{G}^{\sigma'}_{j\wv' + \qv,q\wv'}- \delta_{\sigma\sigma'}\mathcal{G}^\sigma_{i\wv, q\wv' }\mathcal{G}^{\sigma'}_{j\wv' + \qv,p\wv+\qv}),
\end{aligned}
\end{equation}
where $M$ is the number of basis functions per $\wv$ point, $\mathcal{T}^{\sigma\wv}_{ip}$ is the half-rotated one-electron integral matrix and $\psi_{i\sigma}^{\wv}(\pos_P)$ is the half-rotated orthogonal basis function, defined by
\begin{equation}
    \mathcal{T}^{\sigma\wv}_{ip} = \sum_{q} [\Psi_T^{\sigma\dagger}]^\wv_{iq}h_{qp}^\wv,
    \label{roth1}
\end{equation}
and
\begin{equation}
    \psi_{i\sigma}^\wv(\pos_P) = \sum_p [\Psi_T^{\sigma}]^\wv_{pi}\psi_{p}^\wv(\pos_P),
\end{equation}
respectively. Here, we omit the walker index for clarity.
The bottleneck for the local energy evaluation is the calculation of the exchange part given by
\begin{equation}
\begin{aligned}
    E_X = &-\sum_{\substack{ijpq,\sigma\\\wv\wv'\qv}}\sum_{PQ}\psi^{\wv*}_{i\sigma}(\pos_P)\psi^{\wv+\qv}_{p}(\pos_P)M_{PQ}^\qv \psi_{j\sigma'}^{\wv'+\qv*}(\pos_Q)\psi_{q}^{\wv'}(\pos_Q)\\
    &\mathcal{G}^\sigma_{i\wv, q\wv' }\mathcal{G}^{\sigma}_{j\wv' + \qv,p\wv+\qv}.
\end{aligned}
\end{equation}
There are two algorithms with different asymptotic scalings. The algorithm with the lowest asymptotic scaling requires contraction over all the orbital indices and leaves only the ISDF indices uncontracted, as shown in Algorithm \ref{algorithm1},

\begin{algorithm}[H]
\caption{Algorithm A for Exchange energy contraction}\label{algorithm1}
\begin{algorithmic}[1]
\For {$a$ in range($N_\mathrm{slices}$)}
\For {$b$ in range($N_\mathrm{slices}$)}
\State $\displaystyle T_{P_aQ_b}^{\wv, \wv',\sigma} \gets \sum_{iq}\psi_{i}^{\wv*}(\pos_{P_a})\psi_{q}^{\wv'}(\pos_{Q_b})\mathcal{G}_{i\wv, q\wv'}^\sigma$ 
    \Comment{$\mathcal{O}\left(N_{k}^2N_{\mathrm{ISDF}}^2n_{\mathrm{occ}}\right)$}
    \State $\displaystyle S_{P_aQ_b}^{\wv, \wv',\sigma} \gets \sum_{jp}\psi_{j}^{\wv'*}(\pos_{Q_b})\psi_{p}^{\wv}(\pos_{P_a})\mathcal{G}^\sigma_{j\wv', p\wv}$ \Comment{$\mathcal{O}\left(N_{k}^2N_{\mathrm{ISDF}}^2n_{\mathrm{occ}}\right)$}
    \State $\displaystyle E_X \mathrel{+}= -\frac{1}{2}\sum_{\wv,\wv',\qv} \sum_{ab}\sum_{P_a, Q_b}T_{P_aQ_b}^{\wv, \wv',\sigma} S_{P_aQ_b}^{\wv+\qv, \wv'+\qv,\sigma} M_{P_aQ_b}^\qv$
           \Comment{$\mathcal{O}\left(N_{k}^2\log{N_k}N_{\mathrm{ISDF}}^2\right)$}
\EndFor
\EndFor
\end{algorithmic}
\end{algorithm}
 The largest intermediate in this algorithm scales as $\mathcal{O}(N_k^2N_{\mathrm{ISDF}}^2)$, which quickly saturates device memory. Consequently, even for moderately sized systems, we must partition the ISDF index into many slices to satisfy memory limits, and this slicing degrades the efficiency of the contraction kernels. In addition, efficient batched FFT kernels are not available yet on GPUs, which means that we have to use naive loops over $k$-points for the last summation. As a result, we do not maintain the $\mathcal{O}(N_k^2\log N_k)$ scaling anymore. However, for systems with a very small number of $k$-points and a very large simulation cell, this algorithm, even without using FFT, is still useful because of the cubic asymptotic scaling in the number of basis functions per $k$-point. We therefore adopt an alternative algorithm with a higher asymptotic cost but a much smaller memory footprint for simulations with dense $k$-mesh, as shown in Algorithm~\ref{algorithm2}.

\begin{figure}[t]
  \centering
  \begin{minipage}[t]{0.48\linewidth}
    \centering
    \begin{overpic}[width=\linewidth]{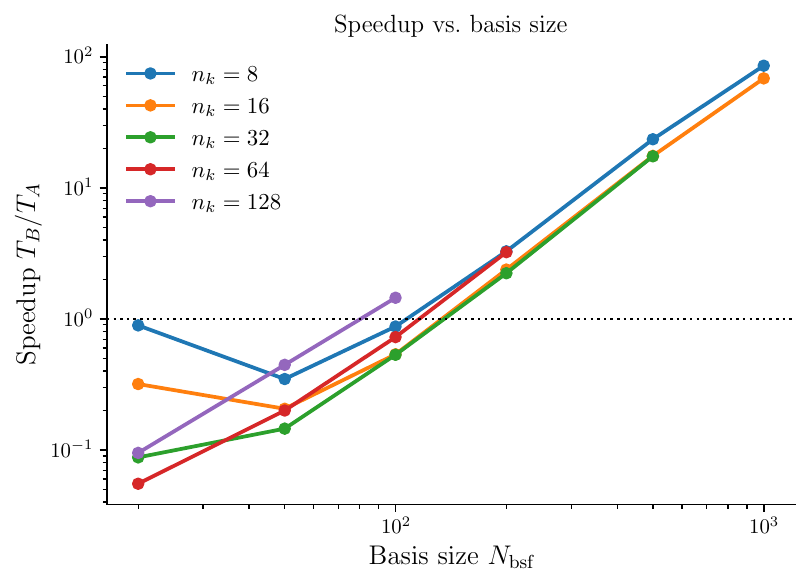}
      \put(2,72){\small (a)}
    \end{overpic}
  \end{minipage}\hfill
  \begin{minipage}[t]{0.48\linewidth}
    \centering
    \begin{overpic}[width=\linewidth]{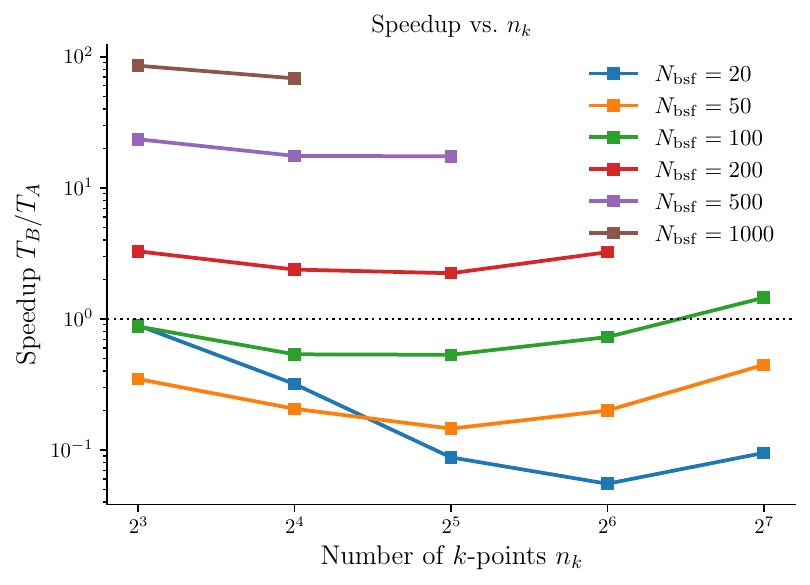}
      \put(2,72){\small (b)}
    \end{overpic}
  \end{minipage}

  \caption{Wall-time comparison of local-energy algorithms A and B across a range of system sizes. All local-energy contractions are implemented via General Matrix Multiplications (GEMMs). Arrays are processed in batches over slices, with the number of slices chosen to fully utilize the available memory capped at 8 GB. The plot shows that, for nearly every $\wv$ mesh, the crossover occurs at roughly 100 basis functions. Below this threshold, algorithm B is generally faster because of its smaller memory footprint. As the number of basis functions increases, however, the superior asymptotic scaling becomes increasingly dominant.}
  \label{fig:alg_comp}
\end{figure}

\begin{algorithm}[H]
\caption{Algorithm B for Exchange energy contraction}\label{algorithm2}
\begin{algorithmic}[1]
\For {$\qv$ in range($N_k$)}
    \State $\displaystyle\rho_{ip}^{\wv, [\qv]}(\pos_P) \gets \psi_{i}^{\wv*}(\pos_P)\psi_{p}^{\wv+[\qv]}(\pos_P)$ 
    \Comment{$\mathcal{O}\left(N_{k}N_{\mathrm{ISDF}}n_{\mathrm{occ}}M\right)$}
    \State $\displaystyle A_{ipQ}^{\wv,[\qv]} \gets \sum_P \rho_{ip}^{\wv, [\qv]}(\pos_P)M^{[\qv]}_{PQ}$ \Comment{$\mathcal{O}\left(N_{k}N_{\mathrm{ISDF}}^2n_{\mathrm{occ}}M\right)$}
    \State $\displaystyle B^{\wv, \wv',[\qv]\sigma}_{pQ} \gets \sum_{j} \psi_{jQ}^{\wv' +[\qv]*}\mathcal{G}_{j\wv' + [\qv], p \wv + [\qv]}^\sigma$
           \Comment{$\mathcal{O}\left(N_{k}^2N_{\mathrm{ISDF}}n_{\mathrm{occ}}M\right)$}
    \State $\displaystyle C^{\wv,\wv', [\qv]}_{iQ} \gets \sum_{p} A_{ipQ}^{\wv,[\qv]}B^{\wv, \wv',[\qv]\sigma}_{pQ}$
           \Comment{$\mathcal{O}\left(N_{k}^2N_{\mathrm{ISDF}}n_{\mathrm{occ}}M\right)$}
    \State $\displaystyle D^{\wv, \wv', [\qv]}_{iq} \gets \sum_Q C^{\wv,\wv', [\qv]}_{iQ} \psi^{\wv'}_{q}(\pos_Q)$ \Comment{$\mathcal{O}\left(N_{k}^2N_{\mathrm{ISDF}}n_{\mathrm{occ}}M\right)$}
    \State $\displaystyle E_X \mathrel{{+}{=}}-\frac{1}{2} \sum_{i, q, \wv, \wv',\sigma}D^{\wv, \wv', [\qv]}_{iq} \mathcal{G}_{i\wv, q\wv'}^\sigma$
\EndFor
\end{algorithmic}
\end{algorithm}
Here $[\qv]$ means that index $\qv$ is held fixed. This algorithm exhibits an overall scaling of $\mathcal{O}\left(N_{k}^2N_{\mathrm{ISDF}}n_{\mathrm{occ}}M(N_k + N_{\mathrm{ISDF}})\right)$, and when using a dense $\wv$ mesh (which is the case in most of the systems we studied in this paper), this algorithm becomes more favorable because of the dominance of the $\mathcal{O}(N_k^2 N^4)$ contribution. Fig. \ref{fig:alg_comp} shows the speedup of Algorithm B against Algorithm A for different system sizes. As shown, Algorithm B is more favorable when the $\wv$-mesh is dense, and the number of basis functions is relatively small. In all the calculations we conducted, we selected the appropriate algorithm for different system sizes to maximize efficiency.

\subsection{THC error in AFQMC calculations}\label{app:thcerror}
\begin{figure}
    \centering
    \includegraphics[width=0.55\linewidth]{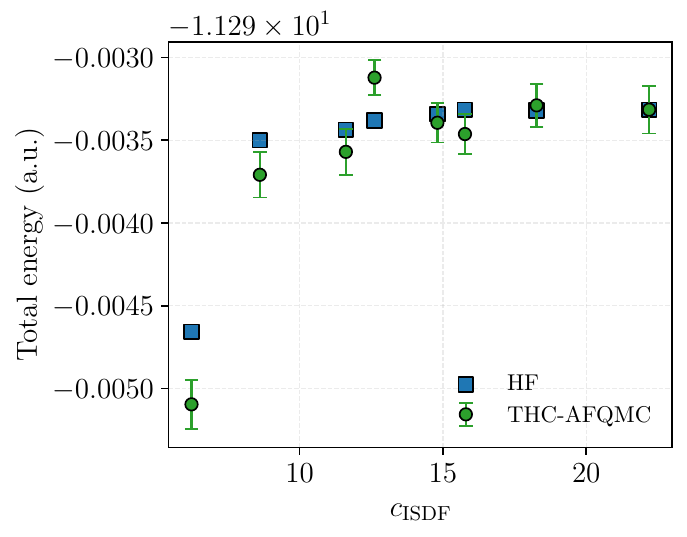}
    \caption{Error in the THC-AFQMC and HF total energies for a $3\times3\times3$ diamond supercell in the DZ basis. For clarity, the HF energies are shifted to match the converged THC-AFQMC total energy.}
    \label{fig:isdferr}
\end{figure}

In this section, we examine the THC-AFQMC error as a function of the number of ISDF grid points. The results are shown in Fig.~\ref{fig:isdferr}. The AFQMC energy converges at $c_{\mathrm{ISDF}} \approx 15$, while HF converges slightly faster with respect to $c_{\mathrm{ISDF}}$. To ensure accuracy in the AFQMC calculations, we select the ISDF grid such that the HF energy error remains below 0.1 mHa throughout this work.

\section{Details for extrapolating to the thermodynamic limit}\label{app:extraptdl}
This section details our extrapolation to the thermodynamic limit, covering (i) test for the linear region for the TDL extrapolation, (ii) time-step extrapolation, and (iii) mean-field twist averaging for metallic systems. TDL linear region tests and time-step tests were validated on diamond. These procedures and conclusions are then applied to the remaining systems whenever needed.

\begin{figure}[htbp]
  \centering
  \begin{minipage}[t]{0.48\textwidth}
    \centering
    \begin{overpic}[width=\linewidth]{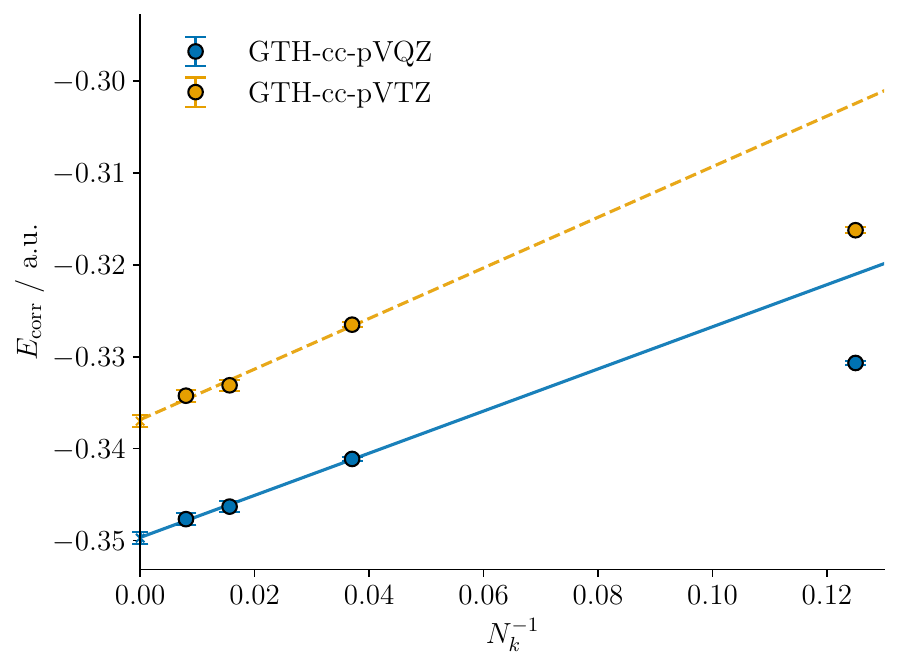}
      \put(3,70){(a)}
    \end{overpic}
    \label{fig:tdlextrap-C}
  \end{minipage}\hfill
  \begin{minipage}[t]{0.48\textwidth}
    \centering
    \begin{overpic}[width=\linewidth]{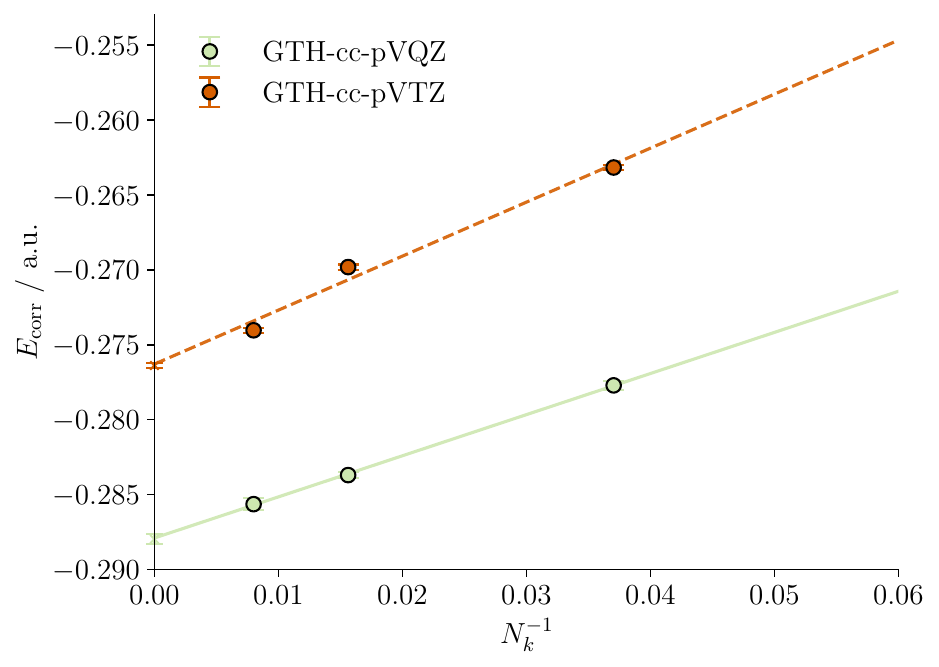}
      \put(3,70){(b)}
    \end{overpic}
    \label{fig:tdlextrap-Si}
  \end{minipage}

  \vspace{0.6em}

  \begin{minipage}[t]{0.48\textwidth}
    \centering
    \begin{overpic}[width=\linewidth]{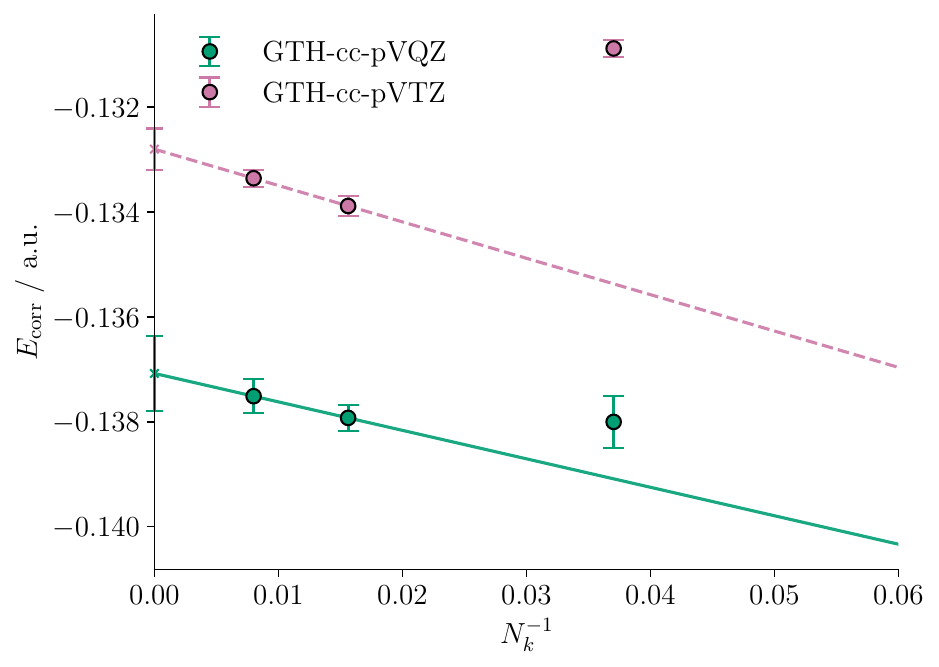}
      \put(3,70){(c)}
    \end{overpic}
    \label{fig:tdlextrap-Li}
  \end{minipage}\hfill
  \begin{minipage}[t]{0.48\textwidth}
    \centering
    \begin{overpic}[width=\linewidth]{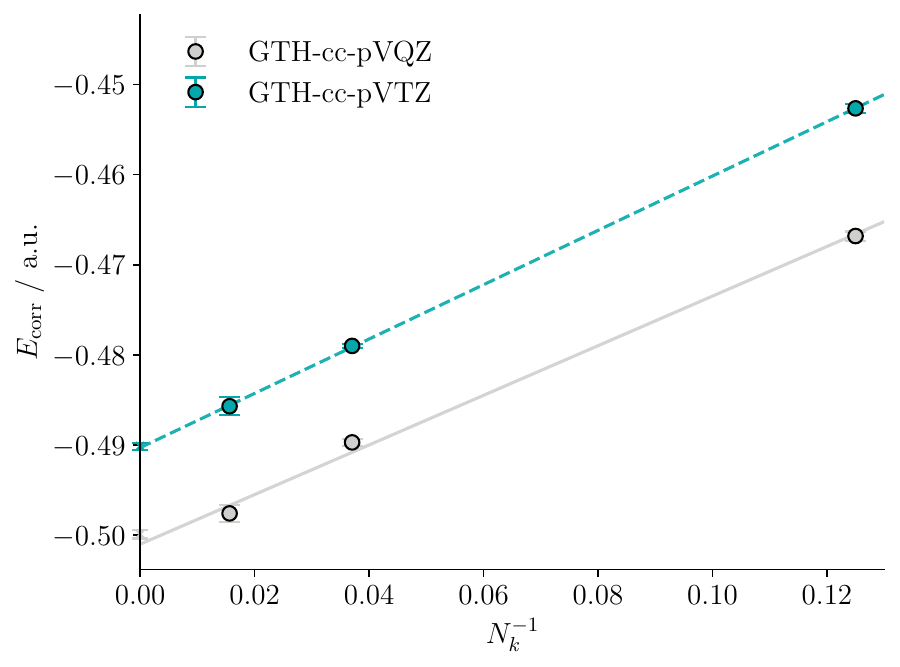}
      \put(3,70){(d)}
    \end{overpic}
    \label{fig:tdlextrap-Al}
  \end{minipage}

  \caption{Extrapolation of the AFQMC correlation energy of (a) diamond; (b) silicon; (c) BCC lithium; (d) FCC aluminum against $N_k^{-1}$ to the thermodynamic limit using the correlation-consistent basis set optimized for the GTH-HF-rev pseudopotential.}
  \label{fig:tdlextrap}
\end{figure}

It has been rigorously proven that, in second-order M{\o}ller-Plesset perturbation theory and periodic coupled-cluster theory, the correlation energy scales inversely with the number of $\wv$ points~\cite{xingUnified2024,xingInverse2024}. Related analyses have also established inverse-volume scaling for diffusion Monte Carlo (DMC) with Jastrow trial wavefunctions~\cite{drummondFinitesize2008,holzmannTheory2016}. Although no analytical expression is currently available for the $\wv$-point scaling of the correlation energy in ph-AFQMC, prior work has reported inverse-volume behavior for sufficiently large $\wv$ meshes~\cite{Malone2020May}. We corroborated this trend by performing AFQMC calculations on $\wv$ meshes ranging from $2\times2\times2$ to $5\times5\times5$, and we observed an approximately linear dependence on $N_k^{-1}$ between $3\times3\times3$ and $5\times5\times5$, as shown in Fig.~\ref{fig:tdlextrap} (a). These meshes correspond to supercells containing 54 to 250 atoms. We therefore adopted this inverse-$N_k$ extrapolation framework for the remaining systems, with system-specific modifications to the $\wv$ meshes and fitting protocol as detailed below.

For silicon, we applied the same protocol as diamond given the same crystal structure. For BCC lithium, since shell effects lead to non-monotonic correlation energies from $3\times3\times3$ to $5\times5\times5$, we adopted a two-point extrapolation using $4\times4\times4$ and $5\times5\times5$, and energy difference between $4\times4\times4$ and $5\times5\times5$ is only approximately 0.5 m$E_h$, which means that the correlation energy is already close to convergence. Hence, the error introduced by the extrapolation is at most 1 m$E_h$. For FCC aluminum, we extrapolated from $2\times2\times2$ to $4\times4\times4$ because the primitive unit cell contains four atoms. For NiO, we performed a two-point extrapolation of the correlation energy for each magnetic state using $3\times3\times3$ and $4\times4\times4$ $\wv$ meshes. Since the $2\times2\times2$ mesh yields an exchange coupling that differs substantially from the values obtained with $3\times3\times3$ through $5\times5\times5$ already at the Hartree-Fock level, we excluded $2\times2\times2$ from the AFQMC thermodynamic-limit extrapolation as well. For CCO, we performed a linear two-point extrapolation using $2\times2\times2$ and $3\times3\times2$ $\wv$ meshes, based on the system size dependence reported in Ref.~\cite{cuiSystematic2022}.

\begin{figure}
    \centering
    \includegraphics[width=0.55\linewidth]{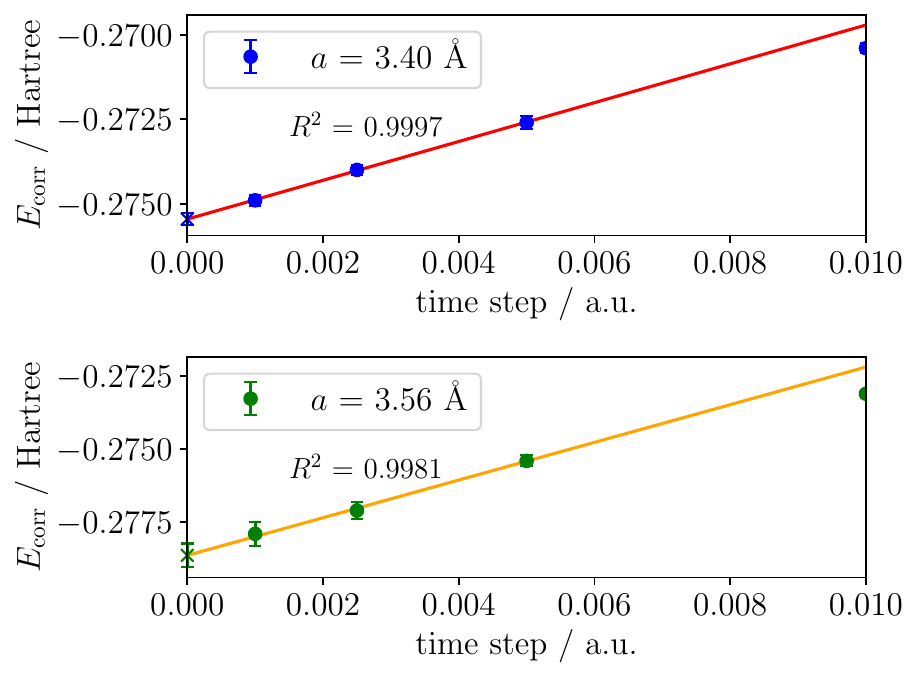}
    \caption{Correlation energy (Hartree) as a function of time step (a.u.) for diamond calculated in GTH-cc-pVDZ basis and $5\times5\times5$ $\wv$-mesh at lattice constants $a = 3.40\ \si{\angstrom}$ and $a = 3.56\ \si{\angstrom}$, respectively.}
    \label{fig:tsextrap}
\end{figure}

Time-step bias in AFQMC depends on the system size, so we used a simple, size-aware protocol to control it without performing full extrapolations at every setting. For each system, we evaluated the observable at two time steps, $\Delta\tau_1=5\times10^{-3}$ a.u. and $\Delta\tau_2=2.5\times10^{-3}$ a.u. Let $E(\Delta\tau_i)$ denote the corresponding estimates with statistical uncertainties $\sigma_i$. If the difference exceeds the combined $1\sigma$ uncertainty, i.e.,
\[
\big|E(\Delta\tau_1)-E(\Delta\tau_2)\big|>\sqrt{\sigma_1^2+\sigma_2^2},
\]
we performed a linear extrapolation in $\Delta\tau$ using these two points and report the $\Delta\tau\!\to\!0$ estimate, which was also reported before by Malone et. al.~\cite{Malone2020May}. As we mentioned in the main text, the reason why we used linear extrapolation is that the size inconsistency error grows linearly with the time step when the time step is small, as shown in Ref.~\cite{leeTwentyYearsAuxiliaryField2022}, and the Trotter and Taylor expansion error grows quadratically~\cite{sukurmaLargeScale2024} with the time step, so they can be neglected. If the two values agree within uncertainties (the $1\sigma$ intervals overlap), we deem the time-step bias to be negligible at the level of our statistical resolution and take the reported value to be the average $\tfrac12\!\left[E(\Delta\tau_1)+E(\Delta\tau_2)\right]$. A preliminary test on diamond (DZ, $5\times5\times5$ $\wv$-mesh) at two lattice constants showed that a time step of $\Delta\tau=10^{-2}$~a.u.\ introduces appreciable nonlinearity in the time-step dependence, as illustrated in Fig.~\ref{fig:tsextrap}. Consequently, we excluded $\Delta\tau=10^{-2}$~a.u.\ from all extrapolations. To avoid prohibitively expensive AFQMC calculations at very small time steps (e.g., $\Delta\tau=10^{-3}$~a.u.), we employed a two-point extrapolation scheme. The resulting fit is indistinguishable, within statistical uncertainty, from that obtained when including the $\Delta\tau=10^{-3}$~a.u.\ data point, thereby validating our two-point extrapolation. Assuming that the time-step error remains linear for $\Delta\tau < 5\times10^{-3}$~a.u., we adopted this extrapolation scheme for all other calculations reported in the main text, which consistently produces smooth TDL extrapolations as expected.

For metallic systems, reaching the thermodynamic limit is more challenging even at the mean field level, since conventional $\Gamma$-centered $\wv$-point sampling often produces oscillatory convergence of the total energy due to the shell effect. To obtain a smooth TDL extrapolation, we employed the twist-averaging technique to mitigate the one-body finite-size error. 
In this work, we used the Baldereschi point as the twist angle (it acts as an ``average'' twist angle), which has proven effective in coupled-cluster studies of simple metals~\cite{Neufeld2022Aug}. We found that applying the twist averaging technique only to the Hartree-Fock energy is sufficient to reach the TDL for the AFQMC energy, i.e., the remaining finite-size effects are fully captured by extrapolating AFQMC correlation energies across different supercell sizes, as shown in Fig.~\ref{fig:tdlextrap}(d).

\section{Details of the error analysis}\label{detailerroranalysis}
\begin{figure}[htbp]
  \centering

  \begin{minipage}[t]{0.4\textwidth}
    \centering
    \begin{overpic}[width=\linewidth]{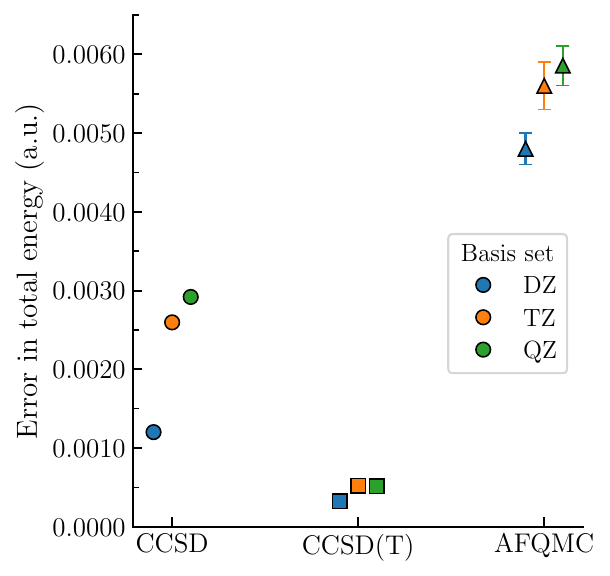}
      \put(3,90){(a)}
    \end{overpic}
    \label{fig:atmerr-C}
  \end{minipage}\hspace{1.8em}
  \begin{minipage}[t]{0.4\textwidth}
    \centering
    \begin{overpic}[width=\linewidth]{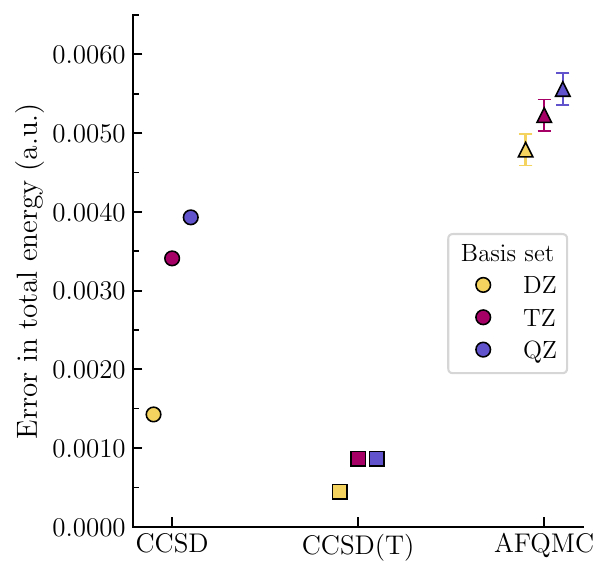}
      \put(3,90){(b)}
    \end{overpic}
    \label{fig:atmerr-Si}
  \end{minipage}

  \vspace{0.6em}

  \begin{minipage}[t]{0.4\textwidth}
    \centering
    \begin{overpic}[width=\linewidth]{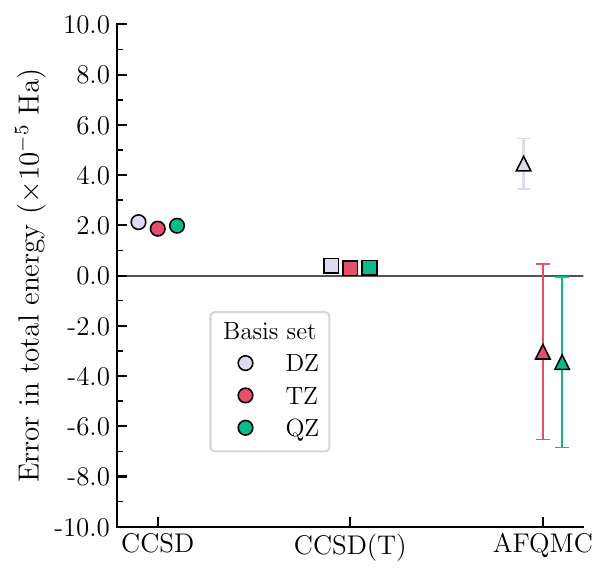}
      \put(3,90){(c)}
    \end{overpic}
    \label{fig:atmerr-Li}
  \end{minipage}\hspace{1.8em}
  \begin{minipage}[t]{0.4\textwidth}
    \centering
    \begin{overpic}[width=\linewidth]{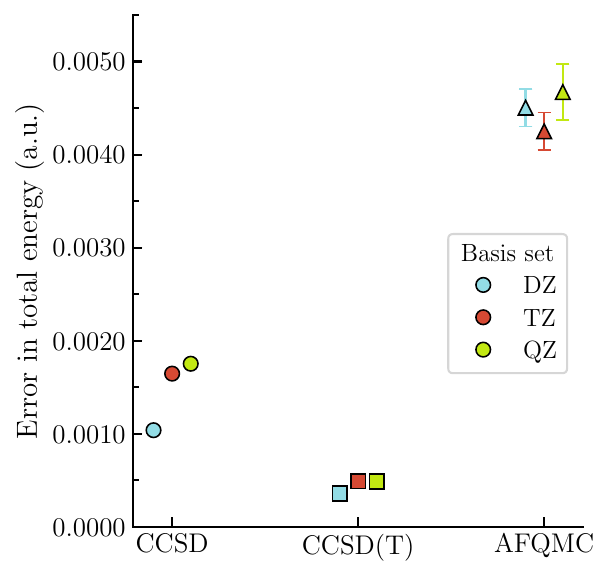}
      \put(3,90){(d)}
    \end{overpic}
    \label{fig:atmerr-Al}
  \end{minipage}

  \caption{Errors of (a) carbon; (b) silicon; (c) lithium; (d) aluminum atomic energies compared to FCI using the correlation-consistent basis set optimized for the GTH-HF-rev pseudopotential. CCSD, CCSD(T) and FCI calculations are performed using \texttt{PySCF}~\cite{sunPySCF2018}; AFQMC calculations are performed with \texttt{ipie}~\cite{maloneIpie2023,jiangImproved2024}.}
  \label{fig:atmerr}
\end{figure}

In this section, we describe the error analysis in detail. In the main text, we classify the total error into two contributions: the phaseless error and the pseudopotential error. For cohesive energies, an additional contribution comes from the atomic phaseless error. Because atoms are small enough to be treated with essentially exact solvers, the atomic phaseless error can be assessed straightforwardly for all the basis sets. The atomic phaseless errors for all systems considered in the main text are shown in Fig.~\ref{fig:atmerr}. These reference results were obtained with \texttt{PySCF}~\cite{sunPySCF2018}. Notably, the Li atom is the only case with a ground-state orbital angular momentum $S$, and it is also the only system with an error below 1 mHa.

Therefore, in the remaining part of the section, we focus on the crystalline phaseless error and the pseudopotential error.

A key complication is that both of these errors depend on system size. Obtaining the exact phaseless error directly at the TDL is infeasible because high-level reference solvers are restricted to relatively small cells. We therefore estimate it by computing the energy difference between AFQMC and a higher-level solver for smaller unit cells and reduced basis sets. We first estimate the basis set dependence on the phaseless error on diamond using only $\Gamma$-point calculation, and we found that the phaseless error is 0.0063(4) Ha for DZ and 0.0070(3) Ha for TZ basis set (the reference result is calculated from FCI). It is thus reasonable to assume that these errors are only weakly dependent on the basis set. Accordingly, we used a double-zeta basis throughout and focus on the system-size dependence of the crystalline phaseless error and the pseudopotential error. 

\begin{figure}
    \centering
    \includegraphics[width=0.8\linewidth]{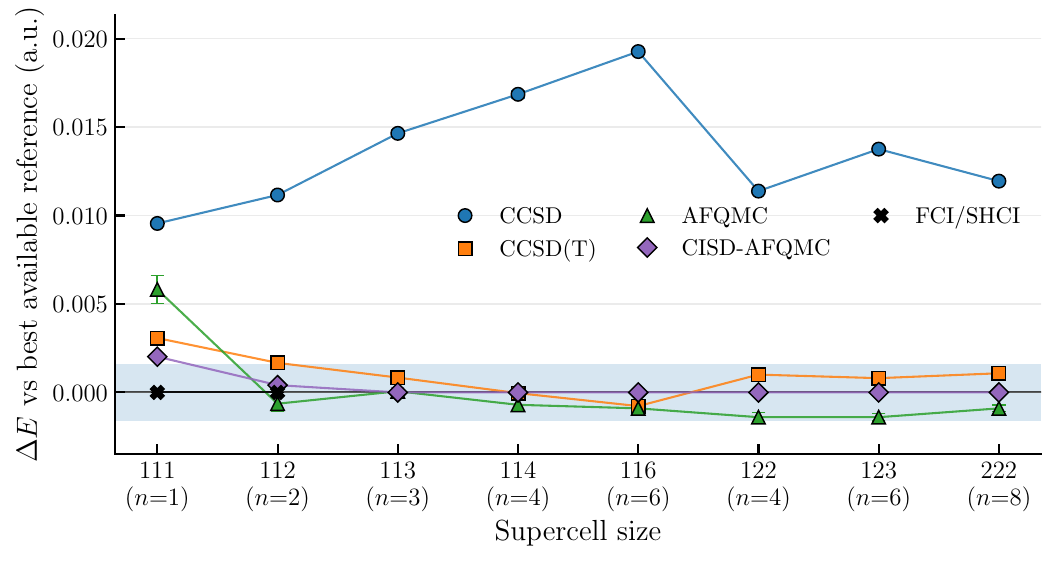}
    \caption{Crystal phaseless error of diamond in the DZ basis evaluated across multiple supercell sizes, where $abc$ denotes a supercell containing $a$, $b$, and $c$ unit cells along the $x$, $y$, and $z$ directions, respectively. For small supercells where essentially exact solvers are feasible, namely FCI or selected CI, we take these as references. Here we used semistochastic heat-bath CI (SHCI)~\cite{sharmaSemistochastic2017} as the selected CI solver which is implemented in \texttt{Dice}~\cite{Dice_github}. For larger supercells beyond the reach of these methods, we used CISD-AFQMC, which is implemented in the \texttt{ad\_afqmc}~\cite{ad_afqmc_github} package, as the reference.}
    \label{fig:c_crysphaseless}
\end{figure}

We first present detailed energy data in DZ basis for diamond across multiple supercell sizes and reference methods, as shown in Fig.~\ref{fig:c_crysphaseless}. CCSD, CCSD(T) and FCI calculations are performed using \texttt{PySCF}~\cite{sunPySCF2018}; Semistochastic heat-bath CI (SHCI)~\cite{sharmaSemistochastic2017} calculations are performed with \texttt{Dice}~\cite{Dice_github}; AFQMC calculations are performed with \texttt{ipie}~\cite{maloneIpie2023,jiangImproved2024}, and CISD-AFQMC calculations are performed with the \texttt{ad\_afqmc}~\cite{ad_afqmc_github} package. For diamond, AFQMC consistently improves upon CCSD, while typically remaining slightly less accurate than CCSD(T). Notably, for supercells with $n \ge 4$, the AFQMC energy drops below the reference, which may indicate mildly nonvariational behavior in larger cells, assuming CISD-AFQMC provides a near-exact benchmark. We therefore used the largest isotropic cell accessible to the higher-level solvers (the $2\times2\times2$ supercell) to estimate and correct the crystalline phaseless error. For other systems, we calculated the corrections on $2\times 2\times2$ supercells. However, the CCSD calculation of Al does not converge on the $2\times 2\times2$ supercell, so we omit the crystalline phaseless error correction for Al. The results are shown in table \ref{tab:crysphls}.
\begin{table}[htbp]
    \centering
    \caption{Crystal phaseless errors for C, Si, Li, and Al in the DZ basis, evaluated on $2\times2\times2$ supercells relative to CISD-AFQMC. Note that these values are not the cohesive-energy corrections directly. The corresponding corrections are obtained by dividing by the number of atoms in the unit cell.}
    \begin{tabular}{ccccc}
    \toprule  
        System & C & Si & Li & Al \\
    \midrule
        Phaseless correction / cell (mHa) & 0.9(2) & 1.0(3) & 2.1(2) & / \\
    \bottomrule
    \end{tabular}
    \label{tab:crysphls}
\end{table}

To assess the pseudopotential error, we compare against cohesive energies from all-electron calculations. Owing to current limitations of the all-electron solver, including severe linear-dependence issues for larger basis sets and supercells, and the absence of an efficient approach to obtain a THC decomposition for all-electron ERIs with $\wv$-point symmetry, the corrections reported here are restricted to DZ calculations on $2\times2\times2$ and $3\times3\times3$ supercells. For $2\times2\times2$ Al, the density-fitting integrals are ill conditioned, leading to large oscillations in the AFQMC energy. We therefore omit the pseudopotential correction for this case. Based on the behavior observed for Si, we nonetheless expect the pseudopotential error for Al to be relatively small. The results are summarized in Table~\ref{tab:allecorr}. Overall, the pseudopotential correction decreases slightly with increasing cell size, but the change from the $2\times2\times2$ to the $3\times3\times3$ supercell is relatively small. We therefore take the $3\times3\times3$ correction as a reliable approximation to the pseudopotential error in the thermodynamic limit.
\begin{table}[htbp]
    \centering
    \caption{All-electron corrections to the cohesive energies of C, Si, Li, and Al in DZ basis for different supercells.}
    \begin{tabular}{cccccc}
    \toprule
       System  &  & C & Si & Li & Al \\
       \midrule
       \multirow{2}{*}{All-electron correction to cohesive energy (mHa)} & $2\times 2\times 2$  & 6.8(3) & 1.9(3)& 7.6(2)& /\\
        & $3\times 3\times 3$ & 5.1(3)& 0.3(3)& 7.47(2) &/\\
    \bottomrule
    \end{tabular}
    \label{tab:allecorr}
\end{table}

\begin{figure}
    \centering
    \includegraphics[width=0.6\linewidth]{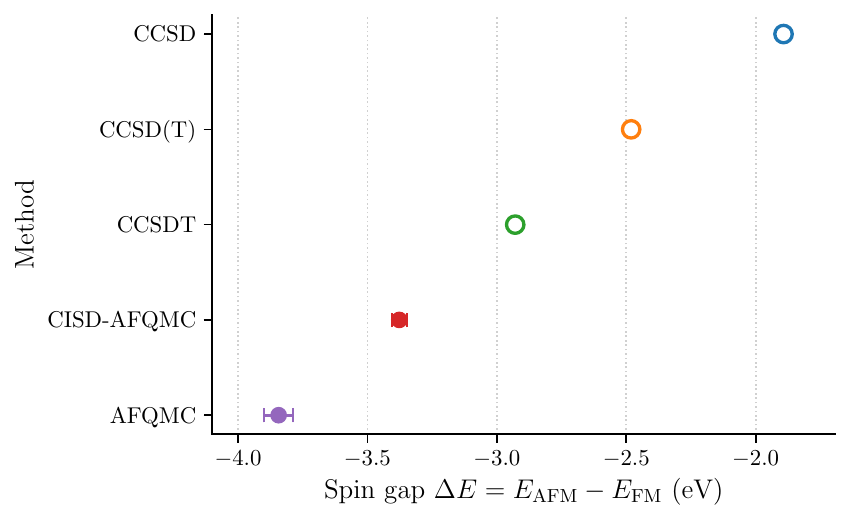}
    \caption{Spin gap $\Delta E = E_\mathrm{AFM} - E_{\mathrm{FM}}$ in eV for $\Gamma$-point NiO in the custom correlation-consistent DZ basis set calculated with various methods. CCSD, CCSD(T) and CCSDT calculations are performed using \texttt{PySCF}~\cite{sunPySCF2018}; CISD-AFQMC calculation is performed with the \texttt{ad\_afqmc}~\cite{ad_afqmc_github} package; AFQMC calculation is performed with \texttt{ipie}~\cite{maloneIpie2023,jiangImproved2024}.}
    \label{fig:niogptbench}
\end{figure}

For NiO, we also assessed the crystalline phaseless error and the pseudopotential error for the Heisenberg $J$ value. While larger supercells or denser k-point sampling would be preferable, the $\Gamma$-point calculation is the only tractable benchmark at this level of theory for NiO. We therefore include $\Gamma$-point NiO results for the sake of completeness, as shown in Fig.~\ref{fig:niogptbench}. However, $\Gamma$-only small-cell benchmarks are not fully representative for AFQMC, whose accuracy typically improves with increasing cell size and or denser $\wv$-point sampling. As shown, AFQMC substantially overestimates the magnitude of the spin gap. Nevertheless, it still outperforms CCSD, assuming CISD-AFQMC is an accurate reference, as supported by its validation for transition-metal oxide molecules in Ref.~\cite{mahajanCCSDT2025}. The pseudopotential error is estimated with all-electron calculation using def2-SVP basis set on $3\times 3\times 3$ supercell, and we obtained $J_2 = -31(2)$ meV for the all-electron calculation, indicating that the pseudopotential error is small.

\section{Matrix Padding for Non-Uniform $\wv$-Point Occupations in Metallic Systems}\label{matpadding} 
In this section, we provide a detailed explanation of how we handle non-uniform $\wv$-point occupations in metallic systems. We denote the number of occupied electrons at each $\wv$-point by $n_{\mathrm{occ}}(\wv)$, the maximum occupation across all $\wv$-points by $n_{\mathrm{occ}}^{\max}$, and the total number of occupied electrons by $n_{\mathrm{occ}}^{\mathrm{tot}}$. For clarity, spin indices are omitted throughout this discussion.

A natural way to address non-uniform $\wv$-point occupations is to adopt a concatenated representation, in which basis functions from all $\wv$-points are assembled into a single block-diagonal structure. In this representation, the trial wavefunction can be expressed as an $(N_k M) \times n_{\mathrm{occ}}^{\mathrm{tot}}$ matrix, obtained by concatenating the occupied orbitals from each $\wv$-point:
\begin{equation}
    \Psi_T = 
    \begin{pmatrix}
      \Psi^{\wv_1} & 0 & \cdots & 0 \\
      0 & \Psi^{\wv_2} & \cdots & 0 \\
      \vdots & \vdots & \ddots & \vdots \\
      0 & 0 & \cdots & \Psi^{\wv_{N_k}}
    \end{pmatrix},
\end{equation}
where each block $\Psi^{\wv_i}$ has dimension $M \times n_{\mathrm{occ}}(\wv_i)$.  
The walker wavefunction in this representation takes the form
\begin{equation}
\Phi = 
\begin{pmatrix}
  \Phi^{\wv_1, \wv_1} & \Phi^{\wv_1, \wv_2} & \cdots & \Phi^{\wv_1, \wv_{N_k}} \\
  \Phi^{\wv_2, \wv_1} & \Phi^{\wv_2, \wv_2} & \cdots & \Phi^{\wv_2, \wv_{N_k}} \\
  \vdots & \vdots & \ddots & \vdots \\
  \Phi^{\wv_{N_k}, \wv_1} & \Phi^{\wv_{N_k}, \wv_2} & \cdots & \Phi^{\wv_{N_k}, \wv_{N_k}}
\end{pmatrix},
\end{equation}
which is generally not block diagonal in $\wv$.

This representation, nonetheless, does not exploit the block-diagonal structure of $\Psi_T$, since the block dimensions are different when the occupation is non-uniform. As a result, one cannot directly use optimized batched matrix-multiplication kernels (e.g. in cuBLAS) to perform the contractions involving the trial wavefunction in this form. To resolve this, we pad each block $\Psi^{\wv_i}$ with zero columns when $n_\mathrm{occ}(\wv_i) < n_{\mathrm{occ}}^{\max}$. The resulting padded trial wavefunction becomes
\begin{equation}
\tilde{\Psi}_T =
\begin{pmatrix}
  \Psi^{\wv_1}\;|\;\mathbf{0} & & & \\
  & \Psi^{\wv_2}\;|\;\mathbf{0} & & \\
  & & \ddots & \\
  & & & \Psi^{\wv_{N_k}}\;|\;\mathbf{0}
\end{pmatrix},
\end{equation}
where the number of zero columns in the $i$th block is $n_\mathrm{occ}^{\max} - n_{\mathrm{occ}}(\wv_i)$.

By permuting columns, this padded matrix can be transformed into
\begin{equation}
    \tilde{\Psi}_T P =
    \begin{array}{c}
    \left(
    \begin{array}{c|c}
      \Psi_T & \mathbf{0}
    \end{array}\right),
    \end{array}
\end{equation}
where $P$ is the corresponding permutation matrix, and the total number of zero columns is $n_\mathrm{occ}^{\max} M - n_\mathrm{occ}^{\mathrm{tot}}$. Applying the same permutation to the walker wavefunction yields $\tilde{\Phi}P = \big(\,\Phi \;\;|\;\; \mathbf{0}\,\big)$. Since zero columns do not affect the nonzero block during matrix multiplications, the propagation step is unchanged.

Difficulties arise, however, when evaluating the padded one-body Green’s function,
\begin{equation}
    \tilde{G} := \left[ \tilde{\Phi} \,(\tilde{\Psi}_T^{\dagger} \tilde{\Phi})^{-1} \tilde{\Psi}_T^{\dagger}\right]^\mathrm{T},
\end{equation}
because the padded overlap matrix $\tilde{S} = \tilde{\Psi}_T^{\dagger} \tilde{\Phi}$ is singular. In fact, one can write
\begin{equation}
    \tilde{S} = 
    P \cdot 
    \begin{pmatrix}
        \Psi_T^\dagger \\ \hline
        \mathbf{0}
    \end{pmatrix}
    \cdot
    \begin{pmatrix}
        \Phi \;\;|\;\; \mathbf{0}
    \end{pmatrix}
    \cdot P^{\dagger}
    =
    P \cdot 
    \left(
\begin{array}{c|c}
S & \mathbf{0} \\ \hline
\mathbf{0} &\mathbf{0}
\end{array}
\right)
    \cdot P^{\dagger},
\end{equation}
where $S = \Psi_T^\dagger \Phi$. Due to the block-diagonal structure, one can safely take the Moore-Penrose pseudoinverse, in which the upper-left block is replaced by the proper inverse of $S$. The resulting padded Green’s function is
\begin{equation}
    \tilde{G} = \left[ \tilde{\Phi} (\tilde{\Psi}_T^{\dagger} \tilde{\Phi})^{+}\tilde{\Psi}_T^{\dagger}\right]^\mathrm{T}
    = \left[ \tilde{\Phi} P \cdot 
    \left(
\begin{array}{c|c}
S^{-1} & \mathbf{0} \\ \hline
\mathbf{0} &\mathbf{0}
\end{array}
\right)
    \cdot P^{\dagger} \tilde{\Psi}_T^{\dagger}\right]^\mathrm{T}
    = \left[\Phi S^{-1}\Psi_T^\dagger\right]^\mathrm{T} = G,
\end{equation}
where $A^+$ denotes the Moore-Penrose pseudoinverse of $A$. We therefore conclude that the matrix padding does not alter the one-body Green’s function. Consequently, the force bias and local energy evaluation also remain unaffected.

\section{Computational details}
\subsection{Semiconductors}
\subsubsection{Diamond}
We used a developer version of \texttt{ipie}~\cite{maloneIpie2023,jiangImproved2024} to carry out all the AFQMC calculations throughout this paper.
AFQMC calculations were carried out on the primitive cell of diamond at the experimental lattice constant ($a = 3.567$ \AA)~\cite{Schimka2011Jan} using a restricted Hartree-Fock (RHF) trial wavefunction from QCPBC~\cite{leeApproaching2021,leeFaster2022,rettigEven2023} (the periodic extension of Q-Chem~\cite{epifanovskySoftware2021}), with all periodic integrals also generated by QCPBC. We used a large-core Goedecker-Teter-Hutter (GTH) pseudopotential optimized for Hartree-Fock calculations (GTH-HF-rev)~\cite{hutterJuerghutter2025} and correlation-consistent basis sets optimized for the corresponding pseudopotential generated by Ye et al.~\cite{yeCorrelationConsistent2022}. The HF energies are converged with respect to the kinetic energy cutoff, which we set as $E_\mathrm{cut} = 600$ eV for diamond. We observed that the walker count needed for a given statistical error decreases as the $k$-point mesh is refined, in agreement with Taheridehkordi et al.~\cite{taheridehkordiPhaseless2023}. After testing, we found that 768 walkers for a $3\times3\times3$ mesh, 384 walkers for $4\times4\times4$, and 192 walkers for $5\times5\times5$ produce negligible population-control bias. As noted by Lee et al.~\cite{leeTwentyYearsAuxiliaryField2022}, AFQMC exhibits size-inconsistency errors at finite time step $\Delta\tau$ that scale linearly for small $\Delta\tau$. We therefore perform two-point extrapolations to the zero time step limit\, using $\Delta\tau={0.005,0.0025}$ for systems larger than TZ/$4\times 4\times 4$ (see Appendix \ref{app:extraptdl}). To reach the thermodynamic limit, we compute correlation energies on $3\times3\times3$, $4\times4\times4$, and $5\times5\times5$ meshes, assuming linear scaling with $N_k^{-1}$. For the CBS limit, we apply Helgaker's two-point extrapolation with an inverse-cubic dependence on the basis-set cardinal number~\cite{helgakerBasisset1997,halkierBasisset1998}. For the largest AFQMC calculation, we used 8 NVIDIA A100 GPU nodes (4 GPUs each node) for 18 h to obtain sufficient statistical samples. The atomic calculations were performed on a single carbon atom with two shells of ghost atoms to remove the BSSE. We used unrestricted Hartree-Fock (UHF) as the trial wavefunction because the ground state of a carbon atom is triplet. Using 768 walkers was sufficient to achieve a statistical error of 0.2 m$E_h$.
\subsubsection{Silicon}
We adopt the experimental lattice constant ($a = 5.43$ Å)~\cite{Schimka2011Jan} and employ the same pseudopotential and basis set as in the diamond calculations. We set the kinetic energy cutoff to $E_\mathrm{cut} = 1000\ \mathrm{eV}$. All other AFQMC settings, including the number of walkers, time step, and extrapolation scheme, are kept identical to those used for diamond. 

\subsection{Metals}\label{app:metals}
For metals, the Hartree-Fock ground state typically exhibits non-uniform $\wv$-point occupations. To address this issue while retaining the efficiency of General Matrix Multiply (GEMM) operations, we adopted the following strategy. The number of occupied electrons is set to the maximum occupation across all $\wv$-points, denoted $n_{\mathrm{occ}}^{\mathrm{max}}$. For $\wv$-points with fewer than $n_{\mathrm{occ}}^{\mathrm{max}}$ occupied states, we padded zero columns so that the final occupation remains uniform across all $\wv$-points. The initial walkers are constructed accordingly, with zero columns added at $\wv$-points of lower occupation. During propagation, these artificial void occupations appear only as zero columns in the walker wavefunctions. They do not affect the physically occupied states, nor do they alter the energy evaluation, since all operations are performed through generalized matrix multiplications. However, the introduction of zero columns renders the overlap matrix between walkers and the trial wavefunction singular, leading to an ill-defined mixed Green’s function. We resolved this by evaluating the Moore-Penrose pseudoinverse of the overlap matrix, which preserves the inverse of the non-zero block and thus restores a well-defined one-body Green's function (see Appendix \ref{matpadding} for more details).  

\begin{figure}[t]
  \centering

  \begin{minipage}{0.46\linewidth}
    \centering
    \begin{overpic}[width=\linewidth]{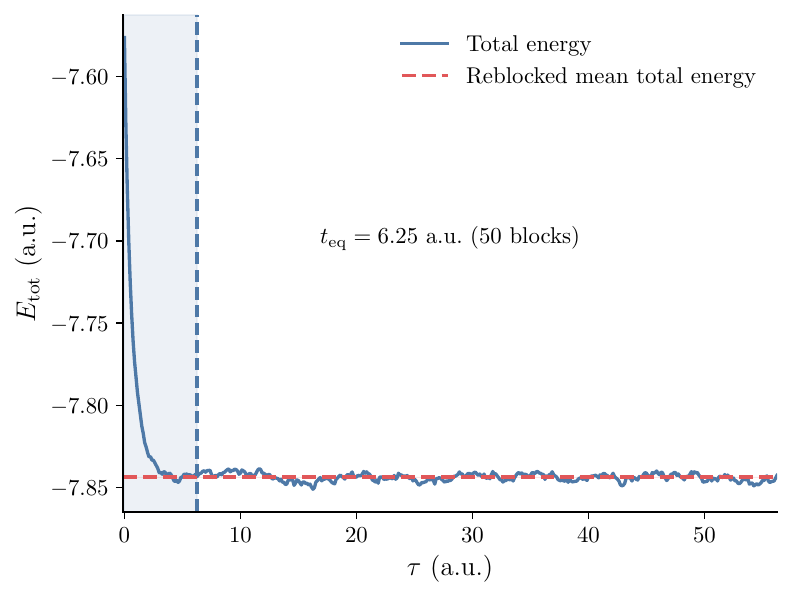}
      \put(2,72){\small{(a)}}
    \end{overpic}
  \end{minipage}\hfill
  \begin{minipage}{0.48\linewidth}
    \centering
    \begin{overpic}[width=\linewidth]{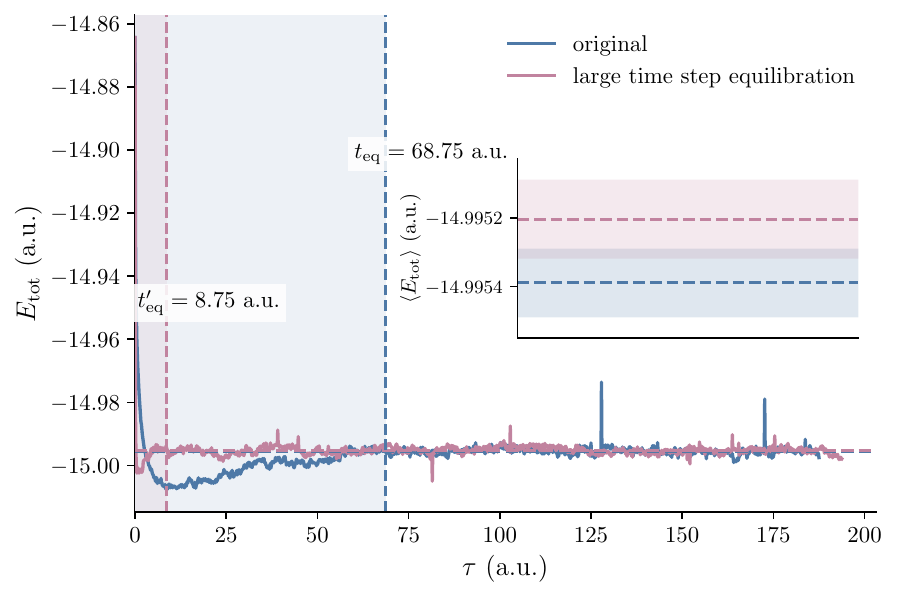}
      \put(2,66){\small(b)}
    \end{overpic}
  \end{minipage}

  \caption{AFQMC total energy trajectories for (a) silicon and (b) BCC lithium using the GTH-cc-pVTZ basis, $4\times4\times4$ $\wv$-point sampling, and time step $\Delta\tau=0.005\ \mathrm{a.u.}$. Panel (a) compares runs with and without a large $\Delta\tau$ pre-equilibration, and the pre-equilibration substantially shortens the equilibration time. The inset of (a) shows reblocked mean energies for both protocols, which agree within error bars, indicating that negligible biases are introduced in this protocol.}
  \label{fig:equilibration_time}
\end{figure}

Another subtlety encountered in the calculation of metallic systems is the unusually long equilibration time as the $\wv$-point mesh becomes denser. For diamond and silicon, equilibration typically requires only < 10 a.u., whereas for lithium with a $4\times 4\times 4$ or $5\times 5\times 5$ $\wv$-mesh the equilibration time can extend to 60-80 a.u., as shown in Fig. \ref{fig:equilibration_time}. We resolved this issue by employing a larger time step during the equilibration phase and subsequently discarding the corresponding weights. The rationale is that although a large time step introduces additional error in the energy estimate, it drives the system much more rapidly toward the equilibrium distribution than a small time step does. With this approach, only $\sim$50 equilibration blocks with large time steps are needed, reducing the total number of equilibration blocks to $\sim$70, compared to $\sim$800 blocks required with the original scheme. We also verified that the results do not change for lithium, as shown in the figure inset. This strategy leads to an order-of-magnitude reduction in equilibration cost, which is especially valuable near the TDL and the CBS limit, where each block can take a long time. We applied this improved equilibration procedure to lithium and aluminum throughout our study, and we expect it to be broadly useful for AFQMC simulations of metallic systems, as well as other systems where the trial wavefunction has a tiny overlap with the true ground state.

\subsubsection{BCC Lithium}
We used the BCC supercell including 2 Li atoms with the experimental lattice constant ($a = 3.453$ \AA)~\cite{Schimka2011Jan} and an RHF trial wavefunction generated with QCPBC. All calculations employ the small-core GTH-HF-rev pseudopotential and the corresponding correlation-consistent basis set generated by Ye \textit{et al.}~\cite{yeCorrelationConsistent2022}. The HF calculations are converged with respect to the kinetic energy cutoff, and we used $E_\mathrm{cut} = 1000$ eV for Li. We used 768, 512, and 384 walkers for the $3\times3\times3$, $4\times4\times4$, and $5\times5\times5$ supercells, respectively. For equilibration, we took $\Delta\tau_{\mathrm{eq}} = 0.05\ \mathrm{a.u.}$ for the $4\times4\times4$ supercell and $\Delta\tau_{\mathrm{eq}} = 0.025\ \mathrm{a.u.}$ for the $5\times5\times5$ supercell. To correct the size-inconsistency error, we performed two-point extrapolations to the zero time-step limit using $\Delta\tau={0.005, 0.0025}$ for systems larger than TZ/$4\times4\times4$. The thermodynamic-limit extrapolation is carried out via a two-point fit using the $4\times4\times4$ and $5\times5\times5$ supercells, as detailed in Appendix~\ref{app:extraptdl}. The Li atom calculations follow the same procedure as for the carbon atom described above.

\subsubsection{FCC Aluminium}
Calculations use the FCC supercell consisting of 4 Al atoms with the experimental lattice constant ($a = 4.018$ \AA)~\cite{Schimka2011Jan} and an RHF trial wavefunction generated with QCPBC. The HF calculations are converged with respect to the kinetic energy cutoff, and we used $E_\mathrm{cut} = 1000$ eV for Al. To mitigate shell effects, the Hartree-Fock reference is extrapolated to the TDL via twist averaging at the Baldereschi angle (Appendix~\ref{app:extraptdl}). We employed the large-core GTH-HF-rev pseudopotential and the corresponding correlation-consistent basis set of Ye \textit{et al.}~\cite{yeCorrelationConsistent2022}. The $2\times2\times2$, $3\times3\times3$, and $4\times4\times4$ supercells use 768, 512, and 384 walkers, respectively. During equilibration we took $\Delta\tau_{\mathrm{eq}} = 0.05\ \mathrm{a.u.}$ for $3\times3\times3$ and $\Delta\tau_{\mathrm{eq}} = 0.025\ \mathrm{a.u.}$ for $4\times4\times4$. For systems larger than TZ/$3\times3\times3$, we performed two-point extrapolations to the zero-time-step limit using $\Delta\tau={0.005, 0.0025}$. The TDL energy is obtained from a three-point extrapolation using the $2\times2\times2$, $3\times3\times3$, and $4\times4\times4$ supercells (Appendix~\ref{app:extraptdl}). The Al atom calculations follow the same protocol as for the carbon atom described above.

\subsection{Transition metal oxides}
The prior GTH-cc-pVXZ basis set work did not include the elements Ca, Ni, and Cu.
Therefore, we generated the corresponding correlation-consistent basis sets for these elements.

\subsubsection{NiO}\label{app:NiO}
We simulated the AFMII and FM phases using a rhombohedral supercell containing 2 formula units (FUs) with an experimental lattice constant $a = 4.17$ \AA (note that this lattice constant is for the cubic supercell of NiO with 4 FUs), and the AFMI phase using a tetragonal supercell containing 2 FUs. Because we found that it is hard to find the AFM solution with an unconstrained HF calculation, the symmetry-broken HF solutions were obtained by performing a HF calculation on the converged SCF for molecular interactions (SCF-MI) calculation. For Ni, we used the small-core GTH pseudopotential optimized for HF to minimize the pseudopotential error (GTH-HF-rev; 18 valence electrons per atom), and for O, we employed the standard large-core GTH-HF-rev pseudopotential. A custom correlation-consistent basis set was optimized for the GTH-HF-rev pseudopotential for Ni, yielding a final basis set consisting of $4s4p3d1f$ shells. Due to linear-dependency issues, $1s1p$ shells with very diffuse exponents were removed from the basis set. The basis set optimization protocol is outlined below: 
\begin{enumerate}
    \item We optimized the HF exponents and contractions as proposed in the original Dunning basis set~\cite{balabanov2005systematically}. As Q-Chem does not natively implement point-group symmetry, we used broken-symmetry restricted-open-shell solutions and the maximum overlap method~\cite{gilbert2008self} to target excited states. The effect of symmetry was found to be negligible when benchmarking for C, for which a correlation-consistent basis set is available.
    \item The correlation exponents were optimized similarly to the original protocol~\cite{balabanov2005systematically} (i.e., state-averaging), but we used CCSD instead of CI methods. 
    Furthermore, we optimized the contraction coefficients directly to lower the correlation energy, rather than taking them from natural orbitals.
    \item We verified the basis sets to see if the Hartree-Fock energy converges to the basis set limit exponentially and the correlation energies to the basis set limit following the $1/Z^3$ trend.
\end{enumerate}

The HF calculations are converged with respect to the kinetic energy cutoff, and we used $E_\mathrm{cut} = 5000$ eV for NiO. we used a timestep of $\Delta \tau = 0.005\ \mathrm{a.u.}$ for all the AFQMC calculations and do not perform time step extrapolation since we expect the time step errors to cancel when calculating the energy differences between magnetic phases. We used 768 and 512 walkers for $3\times 3\times 3$ and $4\times 4\times 4$ calculations, respectively. 

For magnetic moments, the HF and AFQMC results were obtained from a $3\times3\times3$ $\wv$-point simulation using the double-zeta correlation-consistent basis set designed for the GTH-HF-rev pseudopotential.

\subsubsection{\ce{CaCuO2}}\label{app:cco}
We simulated the AFM and FM phases using a unitcell containing 4 FUs with experimental lattice parameters $a = 3.8556$ \AA~and $c = 3.1805$ \AA~\cite{karpinskiSingle1994}. The symmetry-broken HF solutions were obtained by performing an HF calculation on the converged SCF-MI calculation. For Cu, we used the small-core GTH pseudopotential optimized for HF to minimize the pseudopotential error (GTH-HF-rev; 19 valence electrons per atom), for Ca, we used the small-core GTH-HF-rev pseudopotential with 10 valence electrons per atom and for O we employed the standard large-core GTH-HF-rev pseudopotential. A custom correlation-consistent basis set was optimized for the GTH-HF-rev pseudopotential for Cu and Ca, yielding a final basis set consisting of $4s4p3d1f$ and $3s3p2d$ shells, respectively. Due to linear-dependency issues, $1s1p$ shells with very diffuse exponents were removed from the Cu basis set. The basis set exponents and coefficients were optimized using the same protocol as Ni.

The HF calculations are converged with respect to the kinetic energy cutoff, and we used $E_\mathrm{cut} = 3000$ eV for CCO. we used a timestep of $\Delta \tau = 0.005\ \mathrm{a.u.}$ for all the AFQMC calculations.  We used 768 and 512 walkers for $2\times 2\times 2$ and $3\times 3\times 2$ calculations, respectively.
We calculate the Heisenberg $J$ value from AFQMC using a two-point extrapolation scheme between $2\times 2\times 2$ and $3\times 3 \times 2$ $\wv$-meshes. Our HF exchange coupling was obtained from the AFM-FM energy difference in the TDL, determined via a two-point extrapolation using $2\times2\times2$ and $3\times3\times2$ $\wv$-meshes.

\section{Data for the cohesive energies}\label{cohdata}
In this section, we present the cohesive energy data from the various computational methods reported in the main text, along with their corresponding sources. Note that there may be multiple sources of experimental data for some systems. We report the results in the main text from the more recent sources, which have smaller experimental error bars.
\begin{table*}[htbp]
    \centering   
    \renewcommand{\arraystretch}{1.2} 
    \caption{Cohesive energy of diamond predicted by different methods}
    \begin{tabular}{cccc}
        \toprule  
        method & $E_{\mathrm{coh}}$/eV & remark & reference \\    
        \midrule
        HF & 5.17 & / & \multirow{4}{*}{\cite{Ye2024Oct}}\\   
        MP2 & 7.87 & / &  \\
        LNO-CCSD & 7.29 & MP2 TDL/CBS correction & \\
        LNO-CCSD(T) & 7.47 & MP2 TDL/CBS correction & \\
        BWs2 & 7.51 & / & \cite{chenRegularized2025}\\
        DFT/PBE & 7.71 & / & \multirow{2}{*}{\cite{Schimka2011Jan}} \\
        DFT/HSE & 7.61 & / & \\
        DMC & 7.54(1) & B3LYP determinant trial& \cite{Benali2020Nov}\\
        AFQMC & 7.56(1) & TDL extrapolated using $2\times2\times2$ and $3\times 3 \times 3$ k-meshes & \cite{Malone2020May}\\
        AFQMC & 7.53(2)& Direct TDL and CBS extrapolation & \multirow{2}{*}{Our work}\\
        AFQMC$^\dagger$ & 7.50(3)& Corrected for atomic and crystalline phaseless errors and pseudopotential error & \\
        \multirow{2}{*}{Experiment}  & 7.524 & ZPE corrected using the correction from Ref.~\cite{Schimka2011Jan} & \cite{ruscicIntroduction2004}\\
        & 7.545 & ZPE corrected & \cite{Schimka2011Jan}\\
        \bottomrule
    \end{tabular}
    \label{tab:comparison-methods}
\end{table*} 

\begin{table*}[htbp]
    \centering   
    \renewcommand{\arraystretch}{1.2} 
    \caption{Cohesive energy of silicon predicted by different methods}
    \begin{tabular}{cccc}
        \toprule  
        method & $E_{\mathrm{coh}}$/eV & remark & reference \\    
        \midrule
        HF & 2.97 & / & Our work\\   
        MP2 & 4.96 & TDL extrapolated using up to $4\times 4 \times 4$ $\wv$-mesh & \cite{McClain2017Mar} \\
        DFT/PBE & 4.56 & / & \multirow{2}{*}{\cite{Schimka2011Jan}}\\
        DFT/HSE & 4.58 & / & \\
        CCSD & 4.15 & TDL extrapolated using up to $4\times 4 \times 4$ $\wv$-mesh & \cite{McClain2017Mar}\\
        VMC & 4.54(1) & / & \multirow{2}{*}{\cite{leungCalculations1999}}\\
        DMC/LDA & 4.69(1) & / & \\
        DMC/PBE0 & 4.683(3) & Two-point extrapolation using 64 and 216-atom supercells & \cite{Annaberdiyev2021May}\\
        prev. AFQMC & 4.438(3) & Direct TDL extrapolation & \cite{morales2020accelerate}\\
        pw AFQMC & 4.65(3) & Finite-size correction from LDA & \cite{zhangQuantumMonteCarlo2003}\\
        AFQMC & 4.89(2)& Direct TDL and CBS extrapolation & \multirow{2}{*}{Our work}\\
        AFQMC$^\dagger$ & 4.72(3)& Corrected for atomic and crystalline phaseless errors and pseudopotential error & \\
        \multirow{2}{*}{Experiment}  & 4.73 & ZPE corrected using the correction from Ref.~\cite{Schimka2011Jan} & \cite{ruscicIntroduction2004}\\
        & 4.68 & ZPE corrected & \cite{Schimka2011Jan}\\
        \bottomrule
    \end{tabular}
    
    \label{tab:comparison-methods}
\end{table*} 

\begin{table*}[htbp]
    \centering   
    \renewcommand{\arraystretch}{1.2} 
    \caption{Cohesive energy of BCC lithium predicted by different methods}
    \begin{tabular}{cccc}
        \toprule  
        method & $E_{\mathrm{coh}}$/eV & remark & reference \\    
        \midrule
        HF & 0.63 & / & \multirow{3}{*}{\cite{Ye2024Oct}}\\   
        dRPA & 1.21 & / &  \\
        LNO-CCSD & 1.43  & dRPA TDL/CBS correction & \\
        BWs2 & 1.54 & / & \cite{chenRegularized2025}\\
        DMC & 1.56(2) & $N^{-1}$ extrapolation to TDL &\cite{Rasch2015Jul}\\
        VMC & 1.57(1) & LDA finite size correction & \cite{Yao1996Sep}\\
        CCSD & 1.39 & DZ level TDL correction and CBS correction at $3\times 3\times 3$ & \cite{Neufeld2022Aug}\\
        DCSD & 1.50 & Finite size correction using the difference between 16-atom and 8-atom unitcell & \multirow{3}{*}{\cite{Neufeld2023Oct}}\\
        CCSDT & 1.55 & DCSD Finite size correction \\
        DCSDT & 1.55 & DCSD Finite size correction  \\
        DFT/PBE & 1.61 & / &\multirow{2}{*}{\cite{Schimka2011Jan}} \\
        DFT/HSE & 1.57 & / & \\
        AFQMC & 1.45(2)& Direct TDL and CBS extrapolation & \multirow{2}{*}{Our work}\\
        AFQMC$^\dagger$ & 1.69(3)& Corrected for atomic and crystalline phaseless errors and pseudopotential error & \\
        Experiment & 1.66 & ZPE corrected & \cite{Schimka2011Jan} \\
        \bottomrule
    \end{tabular}
    \label{tab:comparison-methods_li}
\end{table*} 

\begin{table*}[htbp]
    \centering   
    \renewcommand{\arraystretch}{1.2} 
    \caption{Cohesive energy of FCC aluminum predicted by different methods}
    \begin{tabular}{cccc}
        \toprule  
        method & $E_{\mathrm{coh}}$/eV & remark & reference \\    
        \midrule
        HF & 1.43 & / & Our work\\  
        BWs2 & 3.67 & / & \cite{chenRegularized2025}\\
        DFT/PBE & 3.43 & / &\multirow{2}{*}{\cite{Schimka2011Jan}} \\
        DFT/HSE & 3.43 & / & \\
        CCSD & 2.97 & DZ level TDL correction and at $3\times 3\times 3$ $\wv$-mesh & \multirow{2}{*}{\cite{Neufeld2022Aug}}\\
        CCSD(T)$_{\mathrm{SR}}$ & 3.10 & DZ level TDL correction and at $3\times 3\times 3$ $\wv$-mesh &  \\
        VMC & 3.23(8) & / & \cite{gaudoinInitio2002}\\
        DMC & 3.403(1) & / & \cite{hoodDiffusion2012}\\
        AFQMC & 3.54(2)& Direct TDL and CBS extrapolation & \multirow{2}{*}{Our work}\\
        AFQMC$^\dagger$ & 3.41(3)& Corrected for atomic and crystalline phaseless errors and pseudopotential error & \\
        Experiment & 3.44 & ZPE corrected & \cite{Schimka2011Jan} \\
        \bottomrule
    \end{tabular}
    \label{tab:comparison-methods_al}
\end{table*}

\section{System sizes in previous AFQMC studies}
Table~\ref{tab:priorafqmc} summarizes the system sizes considered in prior solid-state AFQMC studies. We list the total number of basis functions and electrons, and we briefly describe how finite-size effects were treated for each system.

\begin{table}[htbp]
    \centering
    \caption{Prior solid-state AFQMC studies and the largest system sizes reported. For each work, we list the total number of basis functions and electrons, together with the corresponding finite-size effect treatment. ``PW'' denotes that the entry is the number of plane waves instead of number of molecular orbitals. }
    \begin{tabular}{cccccc}
    \toprule
       System  &  Simulation Cell & $M$ & $N_\mathrm{e}$ & Finite-size effect treatment & Reference\\
       \midrule
       Si & $3\times 3\times 3$ & 5209 (PW) & 216 & Extrapolation & \cite{zhangQuantumMonteCarlo2003}\\
       Si & $3\times 3\times 3$ & Unknown & 216 & Twist averaging over 9 $\wv$-points \& KZK correction & \cite{suewattanaPhaseless2007}\\
       NiO & $1\times 1 \times 1$ & 143 & 48& Twist averaging over $4\times 4\times 4$ $\wv$-mesh \& KZK correction & \cite{maQuantum2015}\\
       Cu & $1\times 1\times 1$ & Unknown & 76 & Twist averaging over $6\times 6\times 6$ $\wv$-mesh \& KZK correction & \multirow{2}{*}{\cite{maAuxiliaryfield2017}}\\
       \ce{H3S} & $ 2\times 2\times 2$ & Unknown & 72 & Twist averaging over $4\times 4\times 4$ $\wv$-mesh \& KZK correction & \\
       NiO & $1\times 1 \times 1$ & 136 & 48& Twist averaging over $4\times 4\times 4$ $\wv$-mesh \& KZK correction & \cite{zhangAuxiliaryfield2018}\\
       MgO & $4\times 4 \times 4$ & 3712 & 1024 & Extrapolation and MP2 correction & \cite{morales2020accelerate} \\
       C & $6\times 6\times 6$ & 7344 & 1728 & Extrapolation & \cite{Malone2020May} \\
       MgO & $4\times 4\times 4$ & 3712 & 1024 & Extrapolation & \cite{zhangAccurate2023} \\
    \bottomrule
    \end{tabular}
    \label{tab:priorafqmc}
\end{table}
\end{document}